\documentclass[a4paper,11pt]{article}
\pdfoutput=1 

\usepackage{jheppub} 
\usepackage{amsmath,amssymb,amsthm,amscd,graphicx}
\usepackage{mathrsfs}
\usepackage[table]{xcolor}                     
\usepackage{colortbl}
\usepackage{booktabs}                          
\usepackage{hyperref}
\usepackage{subfig}	                        
\input epsf.sty

\newcommand\mmm[1]{}

\theoremstyle{definition}

\newcommand{\CA}{{\cal A}}

\newcommand{\CI}{{\cal I}}

\newcommand{\CN}{{\cal N}}
\newcommand{\CO}{{\cal O}}

\newcommand{\CR}{{\cal R}}
\newcommand{\CS}{{\cal S}}

\newcommand{\CV}{{\cal V}}
\newcommand{\CW}{{\cal W}}

\def\IR{{\mathbb R}}

\newcommand{\tr}{{\rm Tr}}
\newcommand{\re}{{\rm e}}
\newcommand{\ri}{{\rm i}}
\newcommand{\rd}{{\rm d}}

\newcommand{\mS}{\mathsf{S}}

\newcommand{\mx}{\mathsf{x}}

\newcommand{\mm}{\mathsf{p}}

\newcommand{\mH}{\mathsf{H}}

\newcommand{\be}{\begin{equation}}
\newcommand{\ee}{\end{equation}}
\newcommand{\ba}{\begin{aligned}}
\newcommand{\ea}{\end{aligned}}
\newcommand{\ben}{\begin{eqnarray}\displaystyle}
\newcommand{\een}{\end{eqnarray}}

\newdimen\tableauside\tableauside=1.0ex
\newdimen\tableaurule\tableaurule=0.4pt
\newdimen\tableaustep
\def\phantomhrule#1{\hbox{\vbox to0pt{\hrule height\tableaurule width#1\vss}}}
\def\phantomvrule#1{\vbox{\hbox to0pt{\vrule width\tableaurule height#1\hss}}}
\def\sqr{\vbox{%
  \phantomhrule\tableaustep
  \hbox{\phantomvrule\tableaustep\kern\tableaustep\phantomvrule\tableaustep}%
  \hbox{\vbox{\phantomhrule\tableauside}\kern-\tableaurule}}}
\def\squares#1{\hbox{\count0=#1\noindent\loop\sqr
  \advance\count0 by-1 \ifnum\count0>0\repeat}}
\def\tableau#1{\vcenter{\offinterlineskip
  \tableaustep=\tableauside\advance\tableaustep by-\tableaurule
  \kern\normallineskip\hbox
    {\kern\normallineskip\vbox
      {\gettableau#1 0 }%
     \kern\normallineskip\kern\tableaurule}%
  \kern\normallineskip\kern\tableaurule}}
\def\gettableau#1{\ifnum#1=0\let\next=\null\else
\squares{#1}\let\next=\gettableau\fi\next}

\newcommand{\bea}{\begin{eqnarray}\displaystyle}
\newcommand{\eea}{\end{eqnarray}}
\newcommand{\nn}{\nonumber}

\newcommand{\mc}[1]{\mathcal{#1}}

\newcommand{\mr}[1]{\mathscr{#1}}
\newcommand{\imag}{\text{Im}}
\newcommand{\real}{\text{Re}}
\newcommand{\jjj}[1]{}
\newcommand{\pd}{\partial}
\newcommand{\sO}{\mathsf{O}}
\newcommand{\cD}{\mathcal{D}}

\newcommand{\wh}[1]{\widehat{#1}}
\newcommand{\wt}[1]{\widetilde{#1}}
\graphicspath{{figs/}}
\newcommand{\figref}[1]{Fig.~\protect\ref{#1}}
\title{{\Huge{\boldmath Thou shalt not tunnel}}\\
\boldmath{Complex instantons and suppression of tunneling in deformed quantum mechanics}}

\author[a]{Jie Gu} 
\author[b]{and Marcos Mari\~no}

\affiliation[a]{School of physics \& Shing-Tung Yau Center\\
  Southeast University, Nanjing 211189, P. R. China}

\affiliation[b]{D\'epartement de Physique Th\'eorique et Section de Math\'ematiques\\
Universit\'e de Gen\`eve, Gen\`eve, CH-1211 Switzerland}

\abstract{The quantization of the Seiberg--Witten curve of
  ${\cal N}=2$ super Yang--Mills theory leads to a deformation of
  one-dimensional quantum mechanics with unconventional behavior. Most
  notably, quantum tunneling is suppressed at special points in
  parameter space. In this paper we examine these deformed models in
  the case of double-well and cubic potentials, and we find that they
  have a rich phase structure. In what we call the strong coupling
  phase, the theory behaves like conventional quantum mechanics,
  instantons are real, and tunneling is not suppressed. In the weak
  coupling phase, the instantons responsible for tunneling become
  complex, and suppression of tunneling takes place at the so-called
  Toda lattice points. At the critical point between the two phases,
  which corresponds to a monopole point in super Yang--Mills theory,
  the non-perturbative amplitudes display an anomalous scaling as a
  function of $\hbar$. This phase structure reflects the physics of
  the underlying super Yang--Mills theory and can be regarded as a
  physical manifestation of wall-crossing behavior of the BPS
  spectrum, which we determine in our problem by using resurgent
  techniques. }

\begin{document}
\maketitle
\flushbottom

\section{Introduction}
\label{sec:intro}

Tunneling through a potential barrier is a quintessential quantum phenomenon with 
no counterpart in classical physics, and its importance 
was recognized from the very beginning of quantum theory \cite{merzbacher}. In addition, tunneling leads to some of the most famous non-perturbative effects in physics, i.e. effects that are exponentially small in $\hbar$ (or the coupling constant). The best known examples are tunneling between two degenerate minima in a double-well potential, which leads to a non-perturbative energy gap, and tunneling in a metastable state, which leads to a non-perturbative imaginary part in the energy of the states (see e.g. \cite{aqm,mmbook} for textbook presentations of these phenomena in one-dimensional quantum mechanics, and \cite{wagner, wagner2} for a recent clarifying discussion.) 

Physicists have been 
also intrigued by the possibility that quantum tunneling be suppressed in certain circumstances, 
and various scenarios where this happens have been proposed. One possibility is to add an external periodic 
field to a situation where tunneling occurs (for example, a double-well potential) 
\cite{cdt-paper}. The Hamiltonian becomes time-dependent and tunneling gets suppressed in 
what has been dubbed coherent destruction of tunneling (CDT). A different mechanism occurs in spin 
systems, where tunneling is suppressed due to the destructive interference between tunneling 
paths \cite{loss,garg}. In particular, in \cite{garg} tunneling is suppressed for a discrete set of values of a control parameter, involving an external magnetic field. More recently, it has been shown that tunneling is suppressed in conventional 
double-well potentials in two dimensions, for certain values of an external magnetic field \cite{fefferman}. 
In the examples involving spin systems, the suppression of tunneling can be understood in 
terms of complex instantons in the coherent path integral for spins, where the complexification is 
due to a Berry phase. Presumably, the suppression observed in \cite{fefferman} can also be
understood in terms of complex instantons, since the external magnetic field introduces a phase 
in the instanton paths (akin to the Aharonov--Bohm effect). Both examples involve tunneling paths in two 
dimensions, either in the two-dimensional sphere describing spin coherent states, or in real space. 
CDT can occur in one dimension, but the instanton picture is less clear due to the presence of a time-dependent Hamiltonian\footnote{In many topological theories anti-instantons, but not instantons, are fully suppressed. This is a very different mechanism from the ones discussed above, see e.g. \cite{fln}.}. 

A different and potentially simpler scenario where quantum tunneling is suppressed 
was discovered in \cite{gm-dqm}. This scenario involves a one-dimensional 
Hamiltonian with an arbitrary polynomial potential of degree $N$, but instead of the conventional non-relativistic kinetic 
term $\mm^2/2$ one has a deformation to $\cosh(\mm)-1$. With such a modified kinetic term, the conventional 
Schr\"odinger differential equation for the wavefunction becomes a finite difference equation. In \cite{gm-dqm} it 
was shown that, when the polynomial potential in the deformed model is the familiar double-well potential, 
tunneling between the two wells is completely suppressed for special values of the parameters in the potential, 
and one finds doubly-degenerate energy states. Tunneling is also suppressed in the 
related example of resonances in an unbounded potential, like e.g. a cubic potential. This is manifested in the 
vanishing of the imaginary part of the energy for certain values of the parameters. 

Why are such deformed quantum-mechanical models interesting? One motivation for their study in \cite{gm-dqm}  
is that they can be regarded as a quantization of the spectral curve of 
the periodic Toda lattice with $N$ particles, or equivalently \cite{mw}, of the Seiberg--Witten 
(SW) curve of ${\cal N}=2$ Yang--Mills theory with gauge group $SU(N)$ \cite{sw}. It is then expected 
that the properties of the deformed quantum mechanics are manifestations of the 
rich properties of these underlying systems. Importantly for us, \cite{gm-dqm} conjectured that 
the suppression of tunneling occurs precisely when the parameters of the polynomial potential 
correspond to the eigenvalues of the quantum periodic Toda lattice. Another motivation for \cite{gm-dqm} 
comes from the TS/ST (topological string/spectral theory) correspondence \cite{ghm,cgm,wzh}, which provides conjectural 
exact quantization conditions (EQCs) 
for many spectral problems based on a quantum spectral curve. The quantization of the SW curve turns out to 
be a relatively simpler case of the correspondence, and \cite{gm-dqm} proposed EQCs in closed form 
for any polynomial potential. These conditions imply in turn the suppression of quantum tunneling at the Toda lattice 
points, and were tested numerically in \cite{gm-dqm}. 

In this paper we would like to provide a better understanding of the
suppression of quantum tunneling in these deformed quantum-mechanical
models by using an instanton approach.  We will focus on the
double-well and cubic potentials, depending on a control parameter,
and we will show that they have two different phases as this parameter
is changed. The phase structure concerns the $\hbar \rightarrow 0$
asymptotics of the non-perturbative effects for the ground state and
the low-lying states in these potentials. The first phase, which we
will call the {\it strong coupling phase}, is qualitatively and
quantitatively very similar to conventional quantum mechanics:
tunneling is never suppressed and is mediated by instantons.  However,
in the so-called {\it weak coupling phase} we have a very different
behavior: instantons become complex and they can interfere
destructively, just as in the spin models of \cite{loss, garg}. This
leads to suppression of tunneling at special points, in agreement with
the picture developed in \cite{gm-dqm}.  There is a critical point
separating the two phases, in which the typical non-perturbative
effects induced by tunneling (energy gap and imaginary part of the
energy) have an ``anomalous" scaling with $\hbar$, which we determine
analytically.

As we mentioned before, one fascinating aspect of the deformed quantum-mechanical models is that its properties 
reflect the rich physics of SW theory. It turns out that 
the existence of two different phases can be understood as a consequence of wall--crossing in the 
BPS spectrum as we go from strong coupling to weak coupling in the SW moduli space 
(this is actually the origin of the names we have given to the phases). In addition, the structure of 
quantum-mechanical instantons corresponds to the spectrum of BPS states in each phase. 
In particular, we cannot obtain the correct non-perturbative effects in the weak-coupling phase by a simple 
analytic continuation of the conventional instantons in the strong coupling phase: we need the new Stokes 
constants, or BPS invariants, associated to the weak coupling phase. We confirm this with a resurgent 
analysis of the perturbative series for the ground state energies in both the double-well and the cubic potential. The 
critical point separating the phases is nothing but a monopole point in the SW moduli space \cite{sw, ds}, where 
$N-1$ monopoles become massless simultaneously, and 
where wall-crossing takes place.  

The paper is organized as follows. In section \ref{sec:dQM} we review
the basic ingredients of the deformed quantum-mechanical models
introduced in \cite{gm-dqm}. In section \ref{sec:strong} we study the
strong coupling phase, which from the point of view of quantum
mechanics corresponds to not too deep potential wells. We generalize
the standard instanton analysis to the deformed case and we compute
the non-perturbative effects at one-loop.  We also show how this
result follows from a WKB-like approach and we test our result against
explicit numerical computations of the spectrum. In section
\ref{sec:weak} we study the weak coupling phase.  We show that
instantons become complex and acquire a phase that leads to
suppression of tunneling, and we note that the naif analytic
continuation of the conventional instantons fails in the case of the
cubic well.  We then explain in detail how a resurgent analysis of the
perturbative series allows one to obtain the correct instanton
structure.  In section \ref{sec:crit} we analyze the theory at the
critical point, and we find that the non-perturbative amplitudes
display an ``anomalous" scaling with $\hbar$, which can however be
understood in detail with the help of the all-orders WKB
method. Finally, in section \ref{sec:con} we conclude and list some
open problems.

\section{Deformed quantum mechanics}
\label{sec:dQM}

The deformed quantum mechanical model we will consider was introduced and studied in \cite{gm-dqm}. It is defined by a 
quantum Hamiltonian of the form
\be
\label{q-ham}
\mH_N=\Lambda^N\left(  \re^\mm+\re^{-\mm} \right) +V_N(\mx), 
\ee
where  
\be
V_N(x)=\sum_{k=0}^{N} c_k \mx^{N-k}
\ee
is a degree $N$ polynomial, and $\mx, \mm$ are the usual Heisenberg operators in $L^2(\IR)$, with 
\be
[\mx, \mm]= \ri \hbar.
\ee
The kinetic term of the Hamiltonian is a deformation of the usual
non-relativistic case $\mm^2/2$. The analogue of the Schr\"odinger
equation is the finite difference equation \footnote{\jjj{Added} It is
  possible to exchange the roles of $x$ and $p$ by doing a Fourier
  transform, and the resulting Hamiltonian operator would involve
  instead a higher order differential operator with a $\cosh$
  potential, instead of a finite difference operator in a polynomial
  potential. However, the instanton picture developed in this paper is
  likely to be more complicated in that Fourier-transformed setting. }
\be
\label{fde}
\Lambda^N \left( \psi(x+ \ri \hbar)+  \psi(x- \ri \hbar) \right) + V_N(x) \psi(x)=E \psi(x).  
\ee
It is sometimes useful to put together the potential and the energy in a single function, 
\be
\label{wnx}
W_N(x)=V_N(x)-E =\sum_{k=0}^{N} (-1)^k h_k \mx^{N-k}.
\ee

Finite-difference equations of the form (\ref{fde}) have been
investigated for a long time. A WKB approach to solve
them was put forward by Dingle and Morgan in \cite{dingle} in the 1960s. More recently they have received a lot of attention, 
since they are related to the quantization of mirror curves and of the SW curves of $\CN=2$
supersymmetric theories. In particular, the Hamiltonian (\ref{q-ham})
was introduced in \cite{gm-dqm} as a direct quantization of the SW
curve of $SU(N)$, $\CN=2$ Yang--Mills theory \cite{sw,klyt}. Most of the recent
approaches to finite-difference equations are based on the WKB method
and its resurgent upgradings, see e.g. \cite{delmonte, baldino,
  hollands-review}.

The approach of \cite{gm-dqm} does not rely on the WKB method and is based on the so-called
TS/ST correspondence \cite{ghm,cgm} (see e.g. \cite{mmrev} for a
review and references). The first aspect of this approach is to regard
$\mH_N$ as an actual operator on $L^2(\IR)$ and study its properties
and spectrum. It is easy to see heuristically, or even
numerically, that, when $N$ is even,  the operator $\mH_N^{-1}$ is trace class, therefore $\mH_N$ 
has a discrete spectrum of energy levels. This has been proved rigorously in
\cite{lst-pol}, building on similar results for the quantization of
mirror curves in \cite{lst}. Although there are no rigorous results
when $N$ is odd, numerical results indicate that in this case the
Hamiltonian $\mH_N$ has a discrete spectrum of resonances, akin to
what is found in conventional quantum mechanics with odd polynomial potentials (see e.g. \cite{aqm}).

Once the spectral problem is well-defined, one can ask whether it can
be solved by explicit EQCs. The main
result of \cite{gm-dqm} was to obtain such conditions as a limiting,
simpler case of the more general EQCs of \cite{ghm,cgm,wzh}. It turns out that 
the EQCs for deformed quantum mechanics 
can be expressed in terms of the so-called Nekrasov--Shatashvili (NS) limit 
\cite{ns} of instanton partition functions \cite{lns,nn}. The Hamiltonian $\mH_N$ is also closely related to the
$Q$-operator of the periodic Toda lattice \cite{gp}. The function
$W_N(x)$ in (\ref{wnx}) can be regarded as a generating function of
the conserved Hamiltonians $h_i$ of this integrable system,
$i=1, \cdots, N$ (the value of $h_0$ sets the overall normalization
and one usually sets $h_{N-1}=0$). The quantum Toda lattice is
characterized by a set of eigenvalues $h_i^{\rm T}$, $i=1, \cdots, N$, for these
Hamiltonians. By using the EQCs of \cite{gm-dqm}, as well as the EQCs
for the Toda lattice conjectured in \cite{ns} and proved in \cite{kt},
one can show the following \cite{gm-dqm}: when the parameters in the
potential $V_N(x)$ of the Hamiltonian (\ref{q-ham}) take the values
corresponding to the Toda lattice eigenvalues, $h_i^{\rm T}$, $i=1, \cdots, N-1$, the energy $E$ is
determined by the value of $h_N^{\rm T}$. In other words, the Toda lattice eigenvalues are special  
solutions of the spectral problem of the Hamiltonian $\mH_N$. For this reason, we will
call these values of the parameters and the energy the {\it Toda
  lattice points}. Note that, when
$N$ is odd, the resonances must become real at these points. In
addition, when $V_4(x)$ has the form of a double-well potential,
\be
V_4(\mx)= h_0\mx^4+ h_2 \mx^2+ h_4,
\ee
one can show that some of the energy levels become degenerate 
at the Toda lattice points. This suggests that quantum tunneling is 
{\it suppressed} at those points, 
in both of its typical manifestations: as a mechanism to lift the degeneracy 
of the spectrum in the double-well potential, and as a mechanism for the decay of metastable states. 

In this paper we would like to gain a better understanding of this suppression of quantum tunneling. To do that, we will 
use instantons, which have become standard tools to explain tunneling and to calculate its effects quantitatively.  Before 
addressing this issue in the next section, let us fix our conventions and state some additional results. 

We will set $\Lambda^N=1/2$, and include an additional constant $-1$ in the potential $V_N(x)$, so that the Hamiltonian will read 
\be \label{eq:HN}
\mH_N= \cosh(\mm) -1 +u_N(\mx). 
\ee
This normalization is chosen in such a way that, for $p$ small, one recovers the standard non-relativistic kinetic term $p^2/2$.  
We will focus on two cases for the potential: the double-well potential for $N=4$, and the cubic potential for $N=3$. The double-well potential will be parametrized as 
\be
\label{u4}
u_4(x)={1\over 2} \left( x^2-{\xi \over 2} \right)^2, \qquad \xi>0. 
\ee
We have chosen the constant term so as to have zero classical energy at the equilibrium points $x=\pm x_\star$, with 
\be
x_\star=  {\sqrt{\xi\over 2}}. 
\ee
In terms of the parameters in $W_4(x)$ we have  
\be
\label{h4}
h_2= -{\xi \over 2}, \qquad h_4= -1+ {\xi^2 \over 8}-E. 
\ee
The cubic potential is taken to be
\be
\label{u3}
u_3(x)={1\over 2} \left( x^3-\xi x+ {2 \xi x_\star \over 3} \right), \qquad x_\star= {\sqrt{\xi \over 3}},  \quad \xi>0,
\ee
so that we also have zero energy at the minimum at $x_\star$. The parameters in $W_3(x)$ are 
\be
\label{h3}
h_2= -{\xi \over 2}, \qquad h_3= -1+ \left( {\xi \over 3} \right)^{3/2}-E. 
\ee
With our normalization, the parameter $\xi$ in the above potentials corresponds to the parameter $h$
used for the quartic and cubic potentials in \cite{gm-dqm} when $\Lambda=1$. 

Let us note that the classical EOM for the Hamiltonian (\ref{eq:HN}) are  
\be
\dot x= \sinh(p), \quad \dot p=-u'(x), 
\ee
so that motion at constant energy $E$ is determined by the equation 
\be
{\sqrt{1+ \dot x^2}}-1+ u(x)= E.  
\ee
This implies that regions where $u(x)>E$ are classically forbidden, as in conventional classical mechanics (when specifying $N$ 
is not necessary, we will write $u(x)$ instead of $u_N(x)$).

\section{Strong coupling phase}
\label{sec:strong}

\subsection{Nonperturbative effects from instantons}
Let us now study instanton configurations in this theory.  Recall
that, in standard quantum mechanics, instantons are responsible for two
important phenomena, as mentioned in the previous section.

The first phenomenon is lifting the degeneracy of the spectrum in the
double-well potential. This energy spectrum 
has naively a double degeneracy at each energy level in
perturbation theory. This degeneracy is broken by quantum tunneling,
and there is a non-perturbatively small difference between $E_{2k}$
and $E_{2k+1}$ for $k=0,1,\ldots$, where $E_{2k}$ and $E_{2k+1}$
correspond respectively to parity even and parity odd
wavefunctions. To detect this small energy difference at the lowest
level, we consider the thermal partition function with anti-periodic
boundary conditions \cite{zj-book}
\begin{equation}
  Z_{\text{a}}(\beta) = \tr \, \mathsf{P}\,  \re^{-\beta \mH/\hbar} = \re^{-\beta E_0/\hbar} -
  \re^{-\beta E_1/\hbar}  + \ldots,
\end{equation}
where $\mathsf{P}$ is the parity operator.
In the semi-classical limit, it is dominated by
\begin{equation}
  Z_{\text{a}}(\beta) \approx \re^{-\beta E_0/\hbar}\frac{\beta}{\hbar}\Delta E +\ldots
\end{equation}
where 
\be
\Delta E = E_1 - E_0 
\ee
is the energy gap between the ground state and the first excited state. We also define the thermal partition function with periodic
boundary conditions,
\begin{equation}
  Z_{\text{p}}(\beta) =\tr \, \re^{-\beta \mH/\hbar} = \re^{-\beta
    E_0/\hbar} + \ldots,
\end{equation}
which in the semi-classical limit is dominated by the ground state
energy. We can express the energy gap $\Delta E$ by
\begin{equation}\label{eq:DelE}
  \Delta E = \lim_{\beta\rightarrow \infty}
  \frac{\hbar}{\beta}\frac{Z_{\text{a}}(\beta)}{Z_{\text{p}}(\beta)}.
\end{equation}
Both $Z_{\text{p}}(\beta)$ and $Z_{\text{a}}(\beta)$ can be computed
by Euclidean path integrals as follows,
\begin{equation}\label{eq:path}
  Z(\beta) = \int \mc{D} x(\tau)\mc{D} p(\tau)
  \exp\left[-\frac{1}{\hbar}
    \int_{-\beta/2}^{+\beta/2}\rd \tau\left(H(x,p)-\ri \dot{x}p\right)
  \right]
\end{equation}
with respectively periodic and anti-periodic boundary
conditions. In the semi-classical limit, $Z_{\text{p}}$ is dominated
by the contribution from the stationary saddle configuration, which we
denote by $Z_0$, while $Z_{\text{a}}$ is dominated by the contribution
from the instanton saddle configuration with anti-periodic boundary
conditions, which we denote by $Z_1^{(-)}$.  Therefore \eqref{eq:DelE}
can also be written as
\begin{equation}
  \Delta E = \lim_{\beta\rightarrow \infty}
  \frac{\hbar}{\beta}\frac{Z_{1}^{(-)}(\beta)}{Z_{0}(\beta)}.
\end{equation}

The second phenomenon is the decay of metastable states in the cubic
potential.  The cubic potential has a discrete spectrum of resonances
which can be obtained by imposing the Gamow-Siegert boundary
condition. The complex energies of these resonant states have
non-perturbatively small negative imaginary parts which characterize
their decay rate. To detect this small imaginary part at the lowest
level, we consider again the thermal partition function with periodic
boundary conditions,
\begin{equation}
  Z(\beta) = \tr \, \re^{-\beta \mH/\hbar} = \re^{-\beta E_0/\hbar} + \ldots.
\end{equation}
This partition function also has a small imaginary part, just like the
energies.  In the semi-classical limit, it is dominated by the ground
state energy, and
\begin{equation}
  E_0 = -\lim_{\beta\rightarrow \infty}\frac{\hbar}{\beta} \log Z(\beta)
\end{equation}
By splitting both $E_0$ and $Z(\beta)$ into a real part and a
imaginary part, one finds that
\begin{equation}
  \imag \, E_0 =
  -\lim_{\beta\rightarrow\infty}\frac{\hbar}{\beta}
  \frac{\imag \, Z(\beta)}{\real \, Z(\beta)}.
\end{equation}
On the other hand, the partition function $Z(\beta)$ is also given by
a path integral formula \eqref{eq:path} with periodic boundary
condition. In the semi-classical limit, it is dominated by the
stationary configuration, and
\begin{equation}
  \real Z(\beta) \approx \real \, Z_0(\beta).
\end{equation}
To find the main contribution to $\imag \, Z(\beta)$, one notices that
the partition function, as a function of $\hbar$, has a branch cut
across the positive real axis, and the discontinuity
\begin{equation}
  Z(\hbar+\ri 0,\beta) - Z(\hbar -\ri 0,\beta) = 2\ri\, \imag \, Z(\hbar,\beta)
\end{equation}
is given at leading order by the contribution from the instanton
configuration, which we denote by $Z_1^{(+)}(\beta)$.
Therefore,
\begin{equation}\label{eq:ImE0}
  \imag \, E_0 =
  -\lim_{\beta\rightarrow\infty}\frac{\hbar}{2\beta}
  \frac{\imag \, Z_1^{(+)}(\beta)}{\real \, Z_0(\beta)}, 
\end{equation}
see \cite{wagner,wagner2} for a recent detailed derivation of this formula, and an illuminating discussion of the relation between 
resonances and instanton calculus. 

To summarize, both the lifting of the ground state energy degeneracy and the decay of
the metastable ground state are characterized by the quantity
\begin{equation}\label{eq:RZ}
  R(\hbar):= \lim_{\beta\rightarrow\infty}\frac{\hbar}{\beta}\frac{Z_1(\beta)}{Z_0(\beta)},
\end{equation}
where $Z_1(\beta)$ is the contribution to the thermal partition
function from the instanton configuration with either periodic or
anti-periodic boundary condition, and $Z_0(\beta)$ is the contribution
from the stationary configuration. These considerations should apply as well to 
deformed quantum mechanics, where the kinetic term in the
Hamiltonian is $\cosh(p)-1$ instead of $p^2/2$ as in \eqref{eq:HN}. We will now verify this in detail.

\subsection{Instanton contribution in deformed quantum mechanics}

Let us consider a generic deformed quantum mechanical model with the
Hamiltonian
\begin{equation}
  H(x,p) = \cosh(p) - 1 + u(x).
\end{equation}
We also assume that the potential $u(x)$ has a prefered local minimum
$x_\star$ where the potential vanishes, i.e.
\begin{equation}
  u(x) \approx \frac{\omega^2}{2}(x-x_\star)^2 + \CO((x-x_\star))^3. 
\end{equation}
In this equation, $\omega$ is the circular frequency.  For instance,
\begin{itemize}
\item the double-well potential \eqref{u4} has two local minima and we
  can choose $x_\star=\sqrt{\xi/2}$ to be the preferred one where
  $\omega = \sqrt{2\xi}$;
\item the cubic potential \eqref{u3} has a preferred local minimum at
  $x_\star = \sqrt{\xi/3}$ where $\omega = (3\xi)^{1/4}$.
\end{itemize}

We are interested in the thermal partition function
\begin{equation}\label{eq:Zbeta}
  Z(\beta) = \int
  \mc{D}x(\tau)\mc{D}p(\tau)\exp\left[-\frac{1}{\hbar}\int_{-\beta/2}^{\beta/2}\rd
    \tau \left(H(x,p) - \ri \dot{x}p\right)\right]
\end{equation}
with either periodic or anti-periodic boundary condition
\begin{equation}
  x(\beta/2) = \pm x(-\beta/2).
\end{equation}
We will consider its dominant contributions in the semi-classical
limit, and also take the low temperature limit $\beta\rightarrow \infty$.

In the semiclassical limit, the thermal partition function is
dominated by the saddle configurations, which are solutions to the
Euclidean EOM. The latter read
\begin{subequations}\label{eq:eom}
  \begin{align}
    &0 = \pd_x H + \ri \dot{p} = u'(x) + \ri \dot{p},\label{eq:eom-x}\\
    &0 = \pd_p H - \ri \dot{x} = \sinh p -\ri \dot{x},\label{eq:eom-p}
  \end{align}
\end{subequations}
from which we find 
the following equation
\begin{equation}\label{eq:Ebeta-t}
  E = 1-\sqrt{1-\dot{x}^2} - u(x).
\end{equation}
Here $E$ can be interpreted as the energy of the saddle configuration,
since in the small $\dot{x}$ limit, it becomes
\begin{equation}
  E \approx \frac{\dot{x}^2}{2} - u(x)
\end{equation}
which is the energy of a conventional mechanical problem with the inverted
potential $-u(x)$.  The location $x_\star$ now becomes a local
maximum.  More precisely, without making the small $\dot{x}$
approximation, \eqref{eq:Ebeta-t} can be recast as
\begin{equation}\label{eq:dyn}
  \frac{\dot{x}^2}{2} - w(x;E) = 0,
\end{equation}
which describes a dynamical system with the ``effective inverted potential''
\begin{equation}
\label{eip}
  -w(x;E) = - u(x)- E + \frac{1}{2}(u(x)+E)^2.
\end{equation}

\begin{figure}
  \centering
  \subfloat[$-u_4$]{\includegraphics[height=5.2cm]{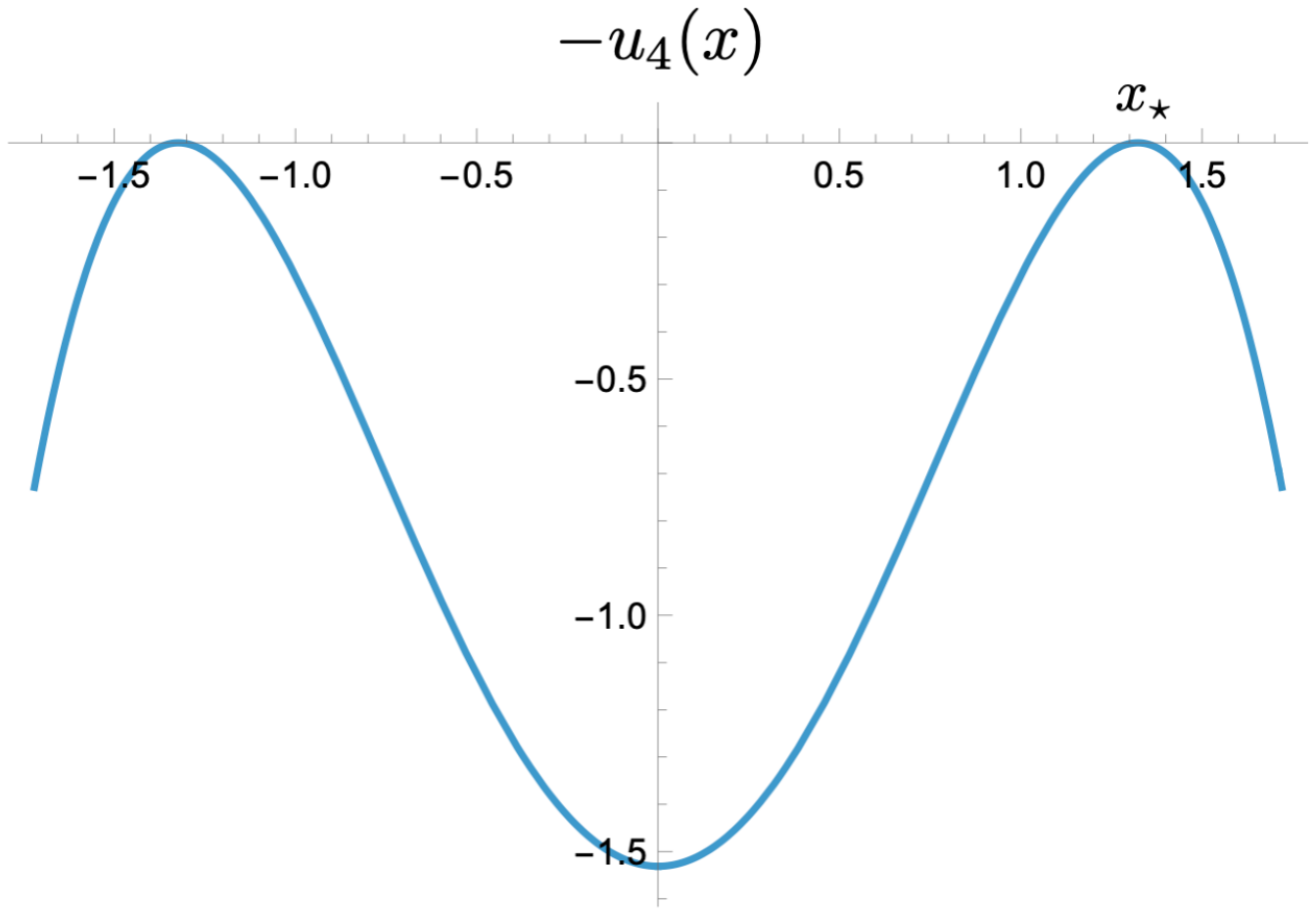}}\hspace{4ex}
  \subfloat[$-w_4$]{\includegraphics[height=5cm]{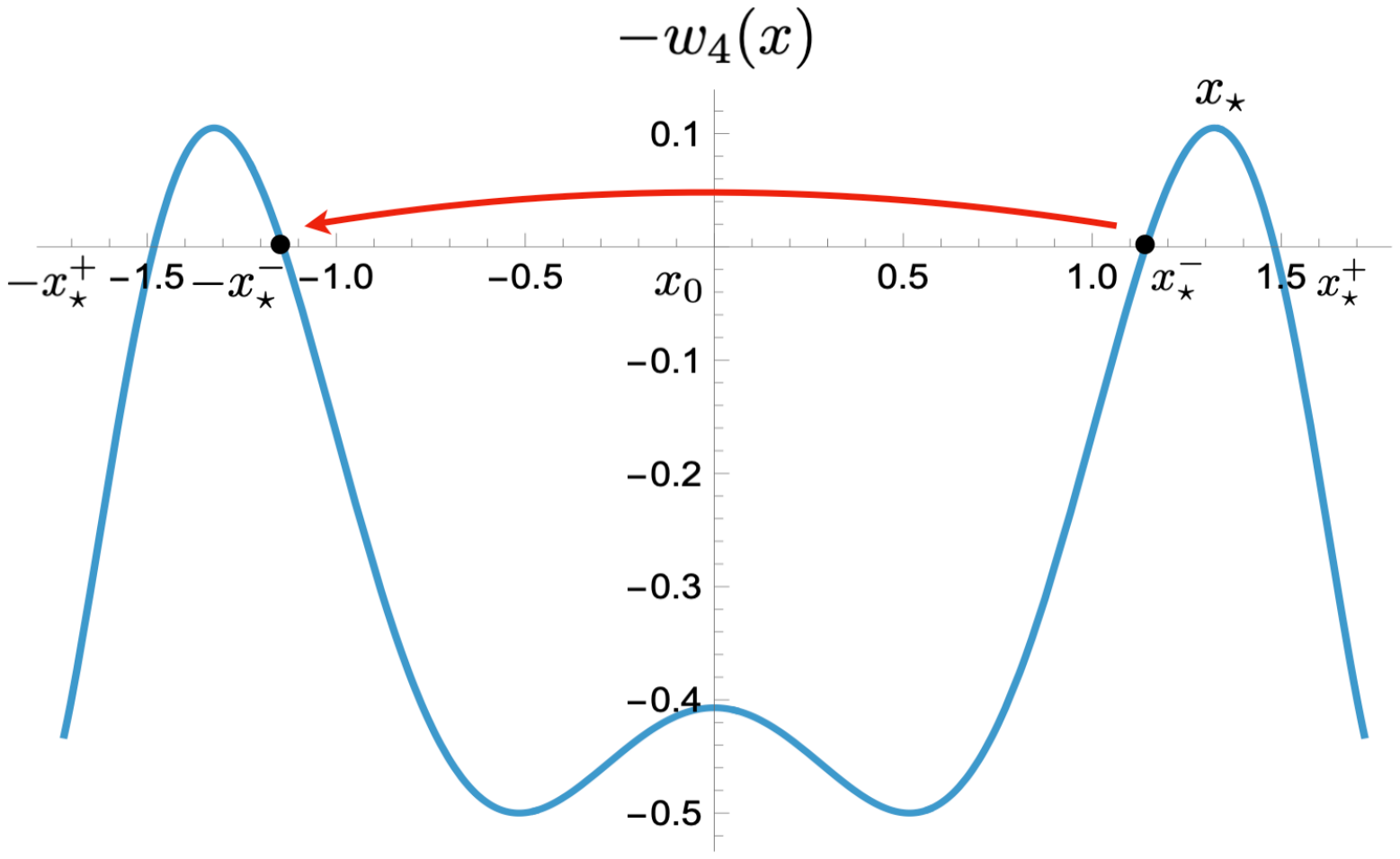}\label{fig:w4dec}}
  \caption{The inverted potentials $-u_4(x)$ and $-w_4(x)$ of the
    double-well model, with $\xi=3.5$ and $E=-0.1$. In the plot of
    $-w_4(x)$ we also indicate the trajectory of the instanton
    configuration.}
  \label{fig:w4}
\end{figure}

\begin{figure}
  \centering
  \subfloat[$-u_3$]{\includegraphics[height=5.2cm]{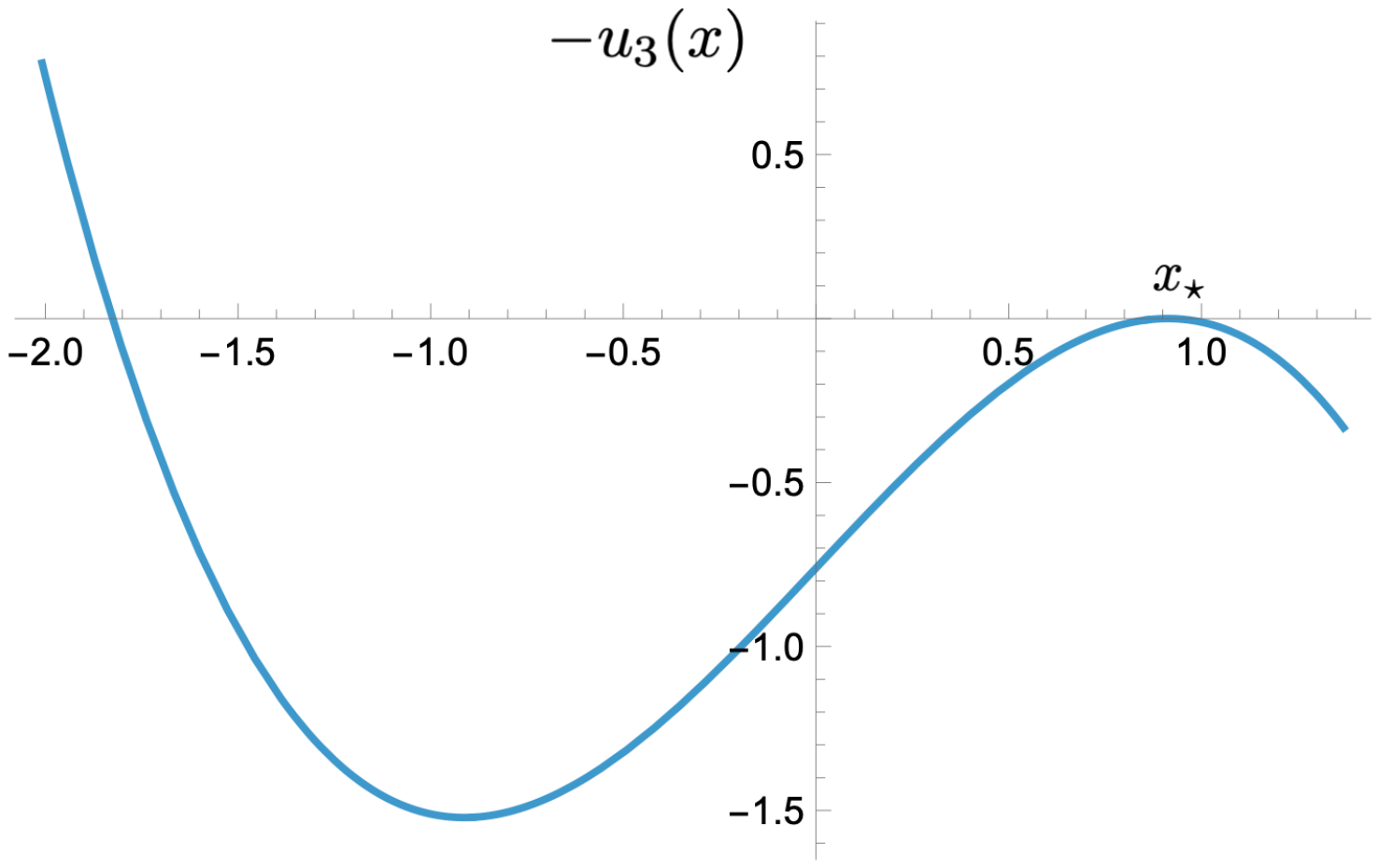}}\hspace{4ex}
  \subfloat[$-w_3$]{\includegraphics[height=5cm]{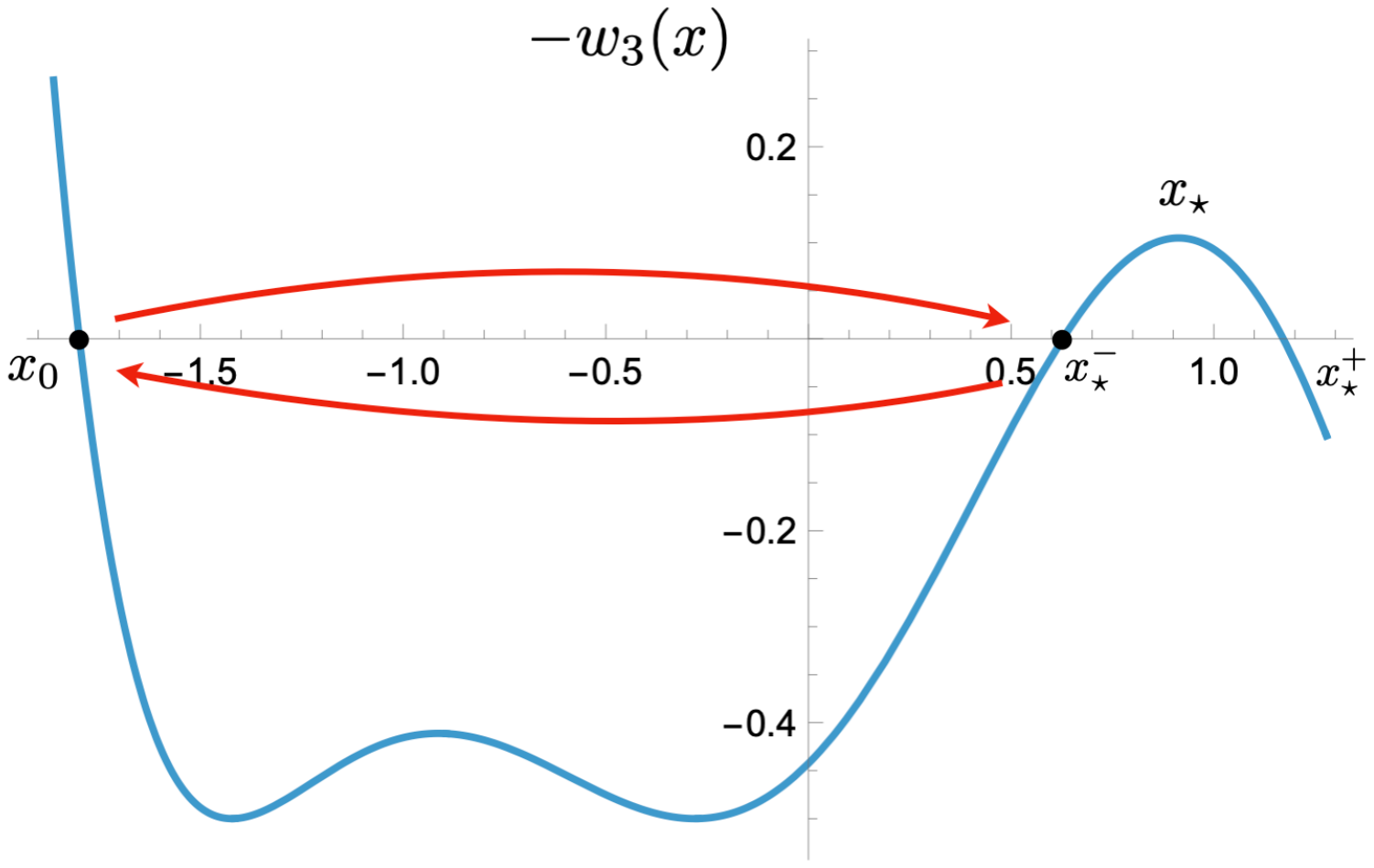}\label{fig:w3dec}}
  \caption{The inverted potentials $-u_3(x)$ and $-w_3(x)$ of the
    cubic model, with $\xi=2.5$ and $E=-0.1$. In the plot of $-w_3(x)$
    we also indicate the trajectory of the instanton configuration.}
  \label{fig:w3}
\end{figure}

\begin{figure}
  \centering
\includegraphics[height=3.2cm]{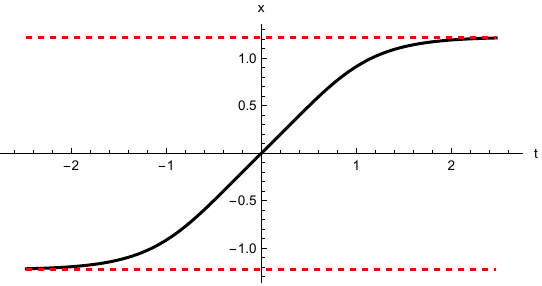}\hspace{4ex}
\includegraphics[height=3cm]{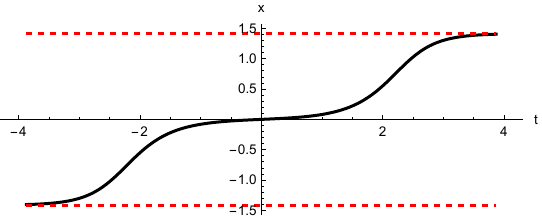}  \caption{The 
trajectories of the instanton $x(\tau)$ in the double well, for $\tau=x_0=0$ and $E=0$. The figure on the left is for $\xi=3$, while 
the figure on the right is for $\xi=3.999$. The horizontal lines denote the asymptotic values $\mp x_\star$.}
  \label{fig:dw-profiles}
\end{figure}
\begin{figure}
  \centering
\includegraphics[height=2cm]{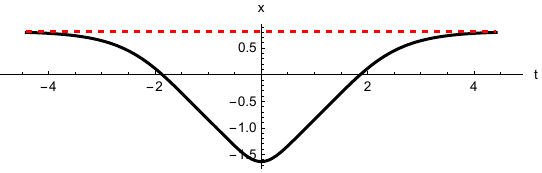}\hspace{4ex}
\includegraphics[height=2cm]{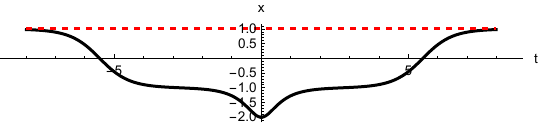}  \caption{The 
trajectories of the instanton $x(\tau)$ in the cubic potential, for $\tau=0$, $x_0$ the turning point, and $E=0$. The figure on the left is for $\xi=2$, while 
the figure on the right is for $\xi=2.999$. The horizontal line is $x_\star$. }
  \label{fig:cw-profiles}
\end{figure}

We consider two solutions to the EOM \eqref{eq:eom}. The first is the
stationary configuration which
stays 
at the local maximum, and it satisfies periodic boundary
conditions. The second is the 1-instanton configuration.  The inverted
potential $-w(x)$ has a potential well next to the local
maximum $x_\star$.  The instanton configuration has negative energy
$E$ and is bounded inside the potential well, with two turning points.
One of the turning points is the solution of $w(x;E) = 0$ which is
near the local maximum $x_\star$ and inside the potential well. We
will denote it by $x_\star^-$.  There are then two different cases
\begin{itemize}
\item In the case of the double-well potential, the instanton
  configuration starts at 
  the turning point $x_\star^-$, and arrives at the opposite turning
  point in the potential well after a time $\beta$, as shown in
  Fig.~\ref{fig:w4}.  It satisfies anti-periodic boundary
  conditions.  Note that due to the parity of the potential, there is also anti-instanton solution 
  whose trajectory is flipped
  horizontally.
\item In the case of the cubic potential, the instanton configuration
  starts from the turning point $x_\star^-$,
  bounces off the opposite turning point in the potential well, and
  returns to $x_\star^-$ after a time $\beta$, as shown in
  Fig.~\ref{fig:w3}. It satisfies periodic boundary conditions.
\end{itemize}
In both cases, we can solve \eqref{eq:Ebeta-t} to obtain $\tau(x)$ as
a function of $x$,
\begin{equation}\label{eq:t-intx}
  \tau -\tau_0 =
  \int_{x_0}^x \frac{\rd x'}{\sqrt{(E+u(x'))(2-E-u(x'))}} \ .
\end{equation}
The instanton configuration has a free parameter or collective coordinate: the time $\tau_0$. At
this time, the instanton is located at certain position $x_0$, that will be
called the center of the instanton and that we will choose conveniently. The instanton trajectories when $E=0$ can be calculated 
explicitly from (\ref{eq:t-intx}) (the integral can be computed in terms of elliptic functions), and we show them in \figref{fig:dw-profiles} and 
\figref{fig:cw-profiles} for the double well and the cubic potential, respectively, and two different values of $\xi$. For small values of $\xi$, these profiles are very similar to the ones found in conventional quantum mechanics, but as we approach the values $\xi=4$ for the double well and $\xi=3$ for the cubic potential, the profiles get deformed and seem to split into pieces. We will come back to this at the end of this section. 


Let us consider the contributions of these two saddle configurations
to the path integral.  First, we consider the stationary solution
which sits at the preferred local maximum $x= x_\star$ of the
inverted potential.  To find its contribution to the path integral, we
expand the Hamiltonian near the stationary solution through
\begin{equation}
  x = x_\star + \delta x,\quad p = \delta p,
\end{equation}
and obtain
\begin{equation}
  H(x,p) = \delta p^2 + \frac{1}{2}\omega^2\delta x^2.
\end{equation}
Its contribution to the partition function reads
\begin{equation}
  Z_0(\beta) \approx \int\mc{D}\delta x \mc{D}\delta p \exp
  \left[-\frac{1}{\hbar}\int_{-\beta/2}^{+\beta/2} \rd \tau\left(\frac{1}{2}\delta
      p^2+\frac{1}{2}\omega^2\delta x^2 - \ri\delta\dot{x}\delta p\right)\right].
\end{equation}
After integrating over $\delta p$, we obtain
\begin{align}
  Z_0(\beta)
  &\approx \int\mc{D}\delta x \exp \left[-\frac{1}{\hbar}
    \int_{-\beta/2}^{+\beta/2}
    \rd \tau \left(\frac{1}{2}\delta\dot{x}^2
    +\frac{1}{2}\omega^2\delta x^2\right)\right]\\
  &= \frac{1}{\sqrt{\det \sO_0}}
    \label{eq:Z0-beta}
\end{align}
where $\sO_0$ is the operator
\begin{equation}\label{eq:O0}
  \sO_0 = -\frac{\pd^2}{\pd\tau^2}+\omega^2.
\end{equation}


Next, we consider the contribution of the instanton configuration,
which we will denote by $x_1(\tau)$ together with $p_1(\tau)$.
Expanding near the instanton configuration,
\begin{equation}
  x = x_1+\delta x,\quad p = p_1 +\delta p,
\end{equation}
we find
\begin{equation}
  Z_1(\beta) =\re^{-\CS/\hbar}\int\cD\delta x\exp\left[
    -\frac{1}{2\hbar}
    \int_{-\beta/2}^{\beta/2}\rd \tau\delta x \tilde{\sO}_1\delta x
  \right]
\end{equation}
where $\CS$ is the action of the instanton configuration
\begin{equation}
  \mc{S} = \int_{-\beta/2}^{\beta/2}\rd \tau (H(x,p)-\ri \dot{x}p),
\end{equation}
and
\begin{equation}\label{eq:wO1}
  \tilde{\sO}_1 = -\sqrt{\cosh p_1}\pd_\tau\frac{1}{\cosh p_1}\pd_\tau\sqrt{\cosh p_1}
  + u''(x_1) \cosh p_1.
\end{equation}
The operator $\tilde{\sO}_1$ has a zero mode (implicitly depending on
$\tau_0$)
\begin{equation}\label{eq:zero}
  \frac{\dot{x}_1(\tau)}{\sqrt{\cosh p_1(\tau)}},
\end{equation}
and after integrating over the zero mode separately, we
obtain (see e.g. \cite{mmbook} for a detailed derivation)
\begin{equation}\label{eq:Z1-beta}
  Z_1(\beta) = \frac{\beta \CW(E)^{1/2}}{\sqrt{2\pi\hbar}}
  \frac{\re^{-\CS/\hbar}}{\sqrt{\det'\tilde{\sO}_1}}.
\end{equation}
where
\begin{equation}\label{eq:tN2}
  \CW(E) = \int_{-\beta/2}^{\beta/2}\rd \tau\frac{\dot{x}_1(\tau)^2}{\cosh p_1(\tau)},
\end{equation}
and $\det'\tilde{\sO}_1$ is the determinant of the operator
$\tilde{\sO}_1$ once the zero mode has been removed.

To evaluate the functional determinants \eqref{eq:Z0-beta} and
\eqref{eq:Z1-beta}, we apply the Gel'fand-Yaglom theorem (see
e.g.~\cite{leon,mmbook,aqm} for an exposition).  Let $\psi_{1,2}$ be
two wavefunctions annihilated by the operator
\begin{equation}
  \sO = -\pd_t^2 + \ldots
\end{equation}
with the boundary conditions
\begin{equation}\label{eq:uj-bdy}
  \begin{aligned}
    \psi_1(-\beta/2) = 1,\quad \dot{\psi}_1(-\beta/2) = 0,\\
    \psi_2(-\beta/2) = 0,\quad \dot{\psi}_2(-\beta/2) = 1. 
  \end{aligned}
\end{equation}
Then, 
\begin{equation}
  \det\sO = \tr\left[
    \begin{pmatrix}
      \psi_1(\beta/2) &\dot{\psi}_1(\beta/2)\\
      \psi_2(\beta/2) &\dot{\psi}_2(\beta/2)\\
    \end{pmatrix}
\mp {\bf 1} \right] = \psi_1(\beta/2) +\dot{\psi}_2(\beta/2) \mp 2,
\end{equation}
where $\mp$ correspond respectively to periodic and anti-periodic
boundary conditions of the wavefunctions in the domain of the operators.

For the operator of the stationary configuration \eqref{eq:O0}, the
two solutions can be chosen to be
\begin{equation}
  \psi_1(\tau) = \cosh \omega(\tau+\beta/2),\quad \psi_2(\tau) =
  \omega^{-1}\sinh \omega(\tau+\beta/2).
\end{equation}
Therefore, the determinant is
\begin{equation}\label{eq:detO0}
  \det\sO_0 = 2\cosh\omega\beta - 2 = 4\sinh^2\omega\beta/2, 
\end{equation}
where we have used that the stationary configuration satisfies
periodc boundary conditions.

To calculate the determinant of the operator of the instanton
configuration \eqref{eq:wO1}, we use the trick that, for an operator
that only has one zero mode, we have \cite{mmbook}
\begin{equation}\label{eq:detp}
  \det\nolimits' \sO = \lim_{\lambda\rightarrow 0}\frac{\rd}{\rd \lambda}\det\sO_\lambda,
\end{equation}
where $\sO_\lambda$ is the perturbed operator
\begin{equation}
  \sO_\lambda = \sO + \lambda.
\end{equation}
The primed determinant in Eq.~\eqref{eq:detp} can also be
calculated by using the Gel'fand-Yaglom theorem.  We consider $\lambda$-dependent 
wavefunctions, annihilated by the perturbed operator $\sO_\lambda$, and we expand them as
\begin{equation}
  \psi_{j,\lambda} = \psi_j^{(0)} + \lambda \psi_j^{(1)} + \ldots
\end{equation}
By solving in power series in $\lambda$, we find that the components must satisfy the following equations
\begin{subequations}\label{eq:O1psi01}
  \begin{align}
    &\sO \psi^{(0)}_j = 0,\\
    &\sO \psi^{(1)}_j = -\psi^{(0)}_j.
  \end{align}
\end{subequations}
Furthermore, the boundary conditions \eqref{eq:uj-bdy} indicate that $\psi_j^{(0)}, \psi_j^{(1)}$ must obey the
following individual boundary conditions
\begin{subequations}\label{eq:uj0-bdy}
  \begin{align}
    \psi_1^{(0)}(-\beta/2) = 1,\quad \dot{\psi}_1^{(0)}(-\beta/2) = 0,\label{eq:uj0-bdy-1}\\
    \psi_2^{(0)}(-\beta/2) = 0,\quad \dot{\psi}_2^{(0)}(-\beta/2) = 1,\label{eq:uj0-bdy-2}
  \end{align}
\end{subequations}
as well as
\begin{subequations}\label{eq:uj1-bdy}
  \begin{align}
    \psi_1^{(1)}(-\beta/2) = 0,\quad \dot{\psi}_1^{(1)}(-\beta/2) = 0,\label{eq:uj1-bdy-1}\\
    \psi_2^{(1)}(-\beta/2) = 0,\quad \dot{\psi}_2^{(1)}(-\beta/2) = 0.\label{eq:uj1-bdy-2}
  \end{align}
\end{subequations}
In terms of these components, it is easy to see that the primed
determinant of $\sO$ is
\begin{equation}\label{eq:detpO1}
  \det\nolimits'\sO = \psi_1^{(1)}(\beta/2) + \dot{\psi}_2^{(1)}(\beta/2).
\end{equation}
Note that this result holds true for both periodic and anti-periodic boundary conditions.

In the case of \eqref{eq:wO1},
we can use the property of the determinant
\begin{equation}
  \det\left(\sO+\lambda\right) = \det(f\cdot\sO\cdot f^{-1}+\lambda),
\end{equation}
to change the normalization of the operator and find
\begin{equation}
  \label{eq:detO1}
  \det\nolimits'\tilde{\sO}_1 = \det\nolimits'\sO_1 =
  \lim_{\lambda\rightarrow 0}\frac{\rd}{\rd \lambda}\det\sO_{1,\lambda}
\end{equation}
where
\begin{equation}\label{eq:O1l}  
  \sO_{1,\lambda} = \sO_1 +\lambda,  
\end{equation}
and
\begin{align}
  \sO_1
  &= \cosh p_1\left(-\pd_{\tau} \frac{1}{\cosh p_1}\pd_{\tau}
    + u''(x_1)\right)\nn\\
  &= - \frac{\pd^2}{\pd \tau^2} +(\pd_{\tau}\log\cosh
    p_1)\frac{\pd}{\pd \tau} + \cosh p_1 u''(x_1).
    \label{eq:O1}
\end{align}
We then proceed as follows: we first find the two solutions $\psi_j^{(0)}$ annihilated by the
operator $\sO_1$.  It is already known from \eqref{eq:zero} that
$\dot{x}_1(\tau)$ is annihilated by $\sO_1$, and is in fact a zero
mode as it satisfies the required periodic or anti-periodic boundary
conditions. It depends implicitly on a free parameter $\tau_0$, and by
choosing it appropriately we can demand
\begin{equation}\label{eq:ddx1}
  \ddot{x}_1(-\beta/2) = 0, 
\end{equation}
which holds e.g. if at $\tau=-\beta/2$ the particle sits at the local minimum of the inverted potential. We can then 
choose $\psi_1^{(0)}$ to be
\begin{equation}
  \psi_1^{(0)}(\tau) = \frac{\dot{x}_1(\tau)}{\dot{x}_1(-\beta/2)}, 
\end{equation}
so that it satisfies the boundary condition
\eqref{eq:uj0-bdy-1}. In addition, as argued in \cite[Sec.~1.6]{mmbook},
\begin{equation}
  \chi(\tau) =\frac{\pd x_1(\tau;E)}{\pd E}
\end{equation}
is also annihilated by $\sO_1$, but it satisfies neither the periodic
nor the anti-periodic boundary conditions.  In fact, by taking
a derivative with respect to $E$ on both sides of \eqref{eq:Ebeta-t}, we
find
\begin{equation}\label{eq:xchi}
  \dot{x}_1(\tau)\dot{\chi}(\tau) - \ddot{x}_1(\tau)\chi(\tau) =
  \cosh p_1(\tau).
\end{equation}
Because of the simplifying assumption \eqref{eq:ddx1} we have made,
\eqref{eq:xchi} implies that
\begin{equation}
  \dot{x}_1(-\beta/2)\dot{\chi}_1(-\beta/2) = \cosh p_1(-\beta/2).
\end{equation}
Therefore, we can choose $\psi_2^{(0)}$ to be
\begin{equation}
  \psi_2^{(0)}(\tau) =
  \frac{\chi(\tau)\dot{x}_1(-\beta/2)-\dot{x}_1(\tau)\chi(-\beta/2)}
  {\cosh p_1(-\beta/2)}, 
\end{equation}
so that it satisfies the boundary
conditions \eqref{eq:uj1-bdy-2}. 
Note that $\psi_j^{(0)}(\tau)$ in addition satisfy another set of
boundary conditions
\begin{equation}\label{eq:psijbdyplus}
  \psi_1^{(0)}(\beta/2) = \dot{\psi}_2^{(0)}(\beta/2) = \pm 1,\quad
  \dot{\psi}_1^{(0)}(\beta/2) = 0,
\end{equation}
thanks to the periodic or anti-periodic boundary conditions of
$x_1(\tau)$.

Next, we propose that $\psi_{j}^{(1)}$ can be constructed as
\begin{equation}
  \label{eq:psij1}
  \psi_j^{(1)}(\tau) = \cosh
  p_1(-\beta/2)\int_{-\beta/2}^{\tau}\rd\tau'
  \frac{\psi_j^{(0)}(\tau')}{\cosh p_1(\tau')}
  \left(\psi_2^{(0)}(\tau)\psi_1^{(0)}(\tau')-\psi_1^{(0)}(\tau)\psi_2^{(0)}(\tau')\right).
\end{equation}
We find that the Wronskian of the solutions $\psi_j^{(0)}$ to
the differential operator \eqref{eq:O1} is
\begin{equation}\label{eq:Wronskian}
  W(\tau) = \psi_1^{(0)}(\tau)\dot{\psi}_2^{(0)}(\tau)
  - \psi_2^{(0)}(\tau)\dot{\psi}_1^{(0)}(\tau) = \frac{\cosh
    p_1(\tau)}{\cosh p_1(-\beta/2)}.
\end{equation}
By using the Wronskian relation, one can easily check that
\eqref{eq:psij1} indeed satisfies the second equation of
\eqref{eq:O1psi01} as well as the boundary conditions
\eqref{eq:uj1-bdy}.  Plugging \eqref{eq:psij1} into \eqref{eq:detpO1} we get
\begin{align}
  \det\nolimits'\sO_1 =
  &\cosh p_1(-\beta/2)
    \left[
    \psi_2^{(0)}(\beta/2)\int_{-\beta/2}^{\beta/2}\rd\tau\frac{\psi_1^{(0)}(\tau)^2}{\cosh
    p_1(\tau)} -
    \dot{\psi}_1^{(0)}(\beta/2)\int_{-\beta/2}^{\beta/2}\rd\tau\frac{\psi_2^{(0)}(\tau)^2}{\cosh
    p_1(\tau)}\right.\nn\\
  &\left.
    + \left(\dot{\psi}_2^{(0)}(\beta/2)-\psi_1^{(0)}(\beta/2)\right)\int_{-\beta/2}^{\beta/2}
    \rd\tau \frac{\psi_1^{(0)}(\tau)\psi_2^{(0)}(\tau)}{\cosh p_1(\tau)}
    \right].
\end{align}
Using the boundary conditions \eqref{eq:psijbdyplus}, it can be
written as
\begin{equation}\label{eq:detpsO1-res0}
   \det\nolimits'\sO_1 = \frac{\chi(\beta/2)\mp
     \chi(-\beta/2)}{\dot{x}_1(-\beta/2)}
   \int_{-\beta/2}^{\beta/2}\rd \tau\frac{\dot{x}_1(\tau)^2}{\cosh p_1(\tau)}.
\end{equation}
Furthermore, taking derivatives with respect to $E$ on the
periodic/anti-periodic boundary condition
\begin{equation}
  x_1(\tau+\beta;E) = \pm x_1(\tau;E)
\end{equation}
we find
\begin{equation}
  \frac{\chi(\beta/2)\mp \chi(-\beta/2)}{\dot{x}_1(-\beta/2)} = \mp
  \frac{\pd \beta}{\pd E},  
\end{equation}
which allows us to further simplify \eqref{eq:detpsO1-res0}
\begin{equation}\label{eq:detpsO1-final}
  \det\nolimits'\sO_1 = \mp \left(\frac{\pd E}{\pd \beta}\right)^{-1}\CW(E),
\end{equation}
where we have also used \eqref{eq:tN2}. Let us note that (\ref{eq:detpsO1-final}) is the same result 
that one obtains in conventional quantum mechanics. The only difference is that the period $\beta$ is computed by 
a different period integral. 

Putting together
\eqref{eq:Z0-beta}, \eqref{eq:detO0}, \eqref{eq:Z1-beta} and \eqref{eq:detpsO1-final},
we conclude that
\begin{equation}\label{eq:Z1oZ0}
  \frac{Z_1(\beta)}{Z_0(\beta)} = \frac{\beta (\CW(E))^{1/2}}{\sqrt{2\pi\hbar}}
  \sqrt{\frac{\det\sO_0}{\det'\tilde{\sO}_1}}\re^{-\CS/\hbar}
  =\frac{2\beta\sinh (\omega\beta/2)}{\sqrt{\mp 2\pi\hbar}} 
  \left(\frac{\pd E}{\pd \beta}\right)^{1/2}\re^{-\CS/\hbar},
\end{equation}
where in the denominator the sign $\mp$ correspond to periodic and
anti-periodic boundary conditions for $Z_1$, respectively.

To proceed
further, we will calculate $\pd E/\pd \beta$. We also note that we have to take the limit $\beta \to \infty$, 
which, as in standard instanton calculus, corresponds to $E \to 0$. By definition,
\begin{equation}\label{eq:betaE}
  \beta(E) = \oint \rd \tau = \oint \frac{\rd x}{\dot{x}} =
  2\int_{x_0(E)}^{x_\star^-(E)} \frac{\rd x}{\sqrt{(E+u(x))(2-E-u(x))}}.
\end{equation}
In the last step, we used the EOM \eqref{eq:Ebeta-t}. $x_0(E)$ is the
centre of the instanton configuration, we can choose it to be $0$ for the
double-well potential (see Fig.~\ref{fig:w4}) and the other turning
point for the cubic potential (see Fig.~\ref{fig:w3}).  Note that this
integral is singular in the limit $E\rightarrow 0$ we are interested in. The function inside the square root factorizes as
\begin{equation}
  (E+u(x))(2-E-u(x)) = (x-x_\star^-(E))(x-x_\star^+(E))p(x;E),
\end{equation}
where $x_\star^{\pm}(E)$ are the two zeros of $u(x)+E$ near the local
maximum $x_\star$.  In the limit $E\rightarrow 0$
\begin{equation}
  x_\star^{\pm}(E) \rightarrow x_\star,
\end{equation}
and the integral becomes
\begin{equation}
  \beta(E) \approx 2\int^{x_\star}\frac{\rd x}{(x_\star-x)\sqrt{p(x;0)}}
\end{equation}
which is divergent.
To regularize this divergence, we write
\begin{align}
\label{split}
  (\omega/2)\beta(E) = 
  &\int_{x_0(E)}^{x_\star^-(E)} \frac{\omega\rd x}{\sqrt{(E+u(x))(2-E-u(x))}}-
    \int_{x_0(E)}^{x_\star^-(E)}
    \frac{\rd x}{\sqrt{(x-x_\star^-(E))(x-x_\star^+(E))}}\nn\\
  &+\int_{x_0(E)}^{x_\star^-(E)}
    \frac{\rd x}{\sqrt{(x-x_\star^-(E))(x-x_\star^+(E))}}.
\end{align}
The first line is now regular as $E\rightarrow 0$, and up to $\CO(E)$
it is given by 
\begin{equation}
\label{calA}
  \CA = \int_{x_0}^{x_\star}\frac{\omega\rd x}{\sqrt{u(x)(2-u(x))}} -
  \int_{x_0}^{x_\star}\frac{\rd x}{x_\star-x},
\end{equation}
where $x_0 = x_0(E=0)$.
The second line is
\begin{equation}
\label{loge}
  \log\frac{\sqrt{x_\star^+(E)-x_0(E)}+\sqrt{x_\star^-(E)-x_0(E)}}
  {\sqrt{x_\star^+(E)-x_0(E)}-\sqrt{x_\star^-(E)-x_0(E)}} \approx
  -\frac{1}{2}\log(-E)+\log \left( \sqrt{2}\omega(x_\star-x_0) \right) + \CO(E).
\end{equation}
Solving $E$ from $\beta$, we arrive at
\begin{equation}
  \frac{\pd E}{\pd \beta} \approx 2\omega^3(x_\star-x_0)^2\re^{-\omega\beta+2\CA}
\end{equation}
as $\beta \rightarrow \infty$. Plugging this into \eqref{eq:Z1oZ0}, we finally conclude that $R(\hbar)$,
defined by \eqref{eq:RZ}, evaluates to
\begin{equation}
  R^{(\pm)}(\hbar) 
  =\sqrt{\frac{\hbar}{\mp\pi}}\omega^{3/2}(x_\star-x_0)\re^{-\CS/\hbar+\CA},
\end{equation}
where the sign $\mp$ in the denominator corresponds to the periodic
and anti-periodic boundary conditions respectively, and $\CS$ is the
instanton action in the $\beta\rightarrow \infty$ limit, or
equivalently, in the $E \rightarrow 0$ limit, which is given by
\begin{equation}
  \mc{S} = \lim_{\beta\rightarrow \infty} \int_{-\beta/2}^{\beta/2}\rd
  \tau (H(x,p)-\ri \dot{x}p) 
  = -\ri \lim_{\beta\rightarrow
    \infty}\oint_{\gamma}  p \rd x = -2\ri
  \int_{x_0}^{x_\star}\cosh^{-1}(1-u(x))\rd x,
\end{equation}
where $\gamma$ is the trajectory of the instanton configuration. It is
also given by 
\be
\label{action-real}
\CS= 2
  \int_{x_0}^{x_\star}\cos^{-1}(1-u(x))\rd x.  
  \ee

We can now write down our final formulae. For the 
double-well potential we find that the energy gap is given by 
\be
\label{egap-ground}
\Delta E= 2R^{(-)}(\hbar) =2  \sqrt{\frac{\hbar}{\pi}} \omega^{3/2} (x_\star - x_0) \re^{-\CS/\hbar+\CA}, 
\ee
with 
\begin{equation}
  x_0 = 0,\;\;  x_\star = \sqrt{\xi/2}, \;\;\omega = (2\xi)^{1/2}.
\end{equation}
In (\ref{egap-ground}) we have included an additional factor of $2$,
which takes into account the contribution of both the instanton and
the anti-instanton. We can also write
\begin{equation}
  \Delta E  =  \sqrt{\frac{\hbar}{\pi}}(2\xi)^{5/4}\re^{-\CS/\hbar+\CA}.
\end{equation}
%
In the case of the cubic potential, the imaginary part of the ground state energy is given by 
\begin{equation}\label{ime-ground}
  \imag \, E = -\frac{1}{2}\imag\, R^{(+)}(\hbar) = -{1\over 2} {\sqrt{\hbar \over \pi}} \omega^{3/2} (x_\star -x_0) \re^{-\CS/\hbar+\CA},
\end{equation}
with
\begin{equation}
  x_0 = -2x_\star, \;\; x_\star = \sqrt{\xi/3},\;\; \omega = (3\xi)^{1/4}.
\end{equation}
We can also write
\be\label{eq:ime-short}
  \imag \, E=-\sqrt{\frac{\hbar}{\pi}} \frac{(3\xi)^{7/8}}{2}\re^{-\CS/\hbar+\CA}. 
\ee
As expected from (\ref{eq:detpsO1-res0}), these formulae are formally
identical to the results obtained in conventional quantum mechanics,
but the quantities $\CS$ and $\CA$ are periods of different
differentials \jjj{Modified}(for example, although $\CS$ is obtained
by integrating the Liouville form $p(x) \rd x $ in both standard
quantum mechanics and deformed quantum mechanics, the expression of
$p(x)$ can be very different.).
 
\begin{figure}
  \centering
  \subfloat[$\xi<\xi_c$]{\includegraphics[height=2.5cm]{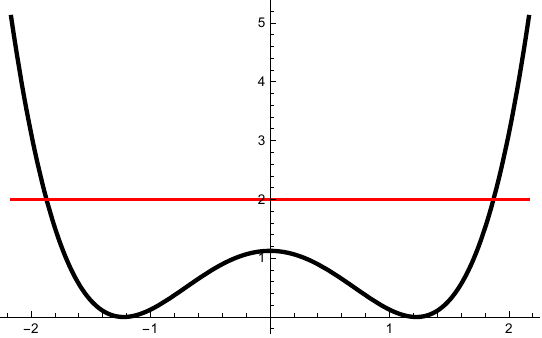}}\hspace{10ex}
  \subfloat[$\xi=\xi_c$]{\includegraphics[height=2.5cm]{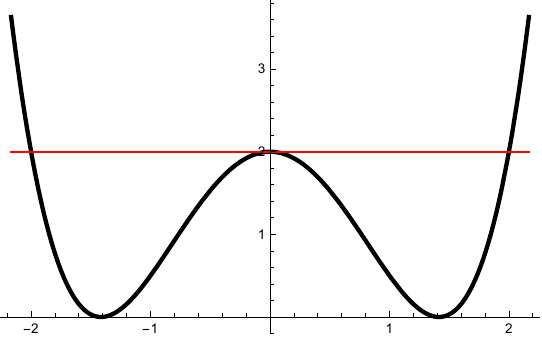}}\hspace{10ex}\subfloat[$\xi>\xi_c$]{\includegraphics[height=2.5cm]{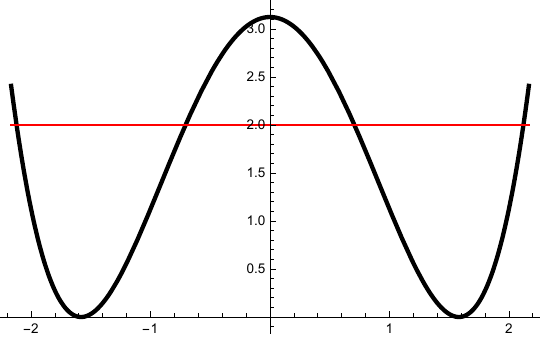}}
  \caption{As we change the value of the parameter $\xi$ in the double well and cubic potentials, the theory undergoes a phase transition at a critical value $\xi=\xi_c$. Here we show the change in the form of the potential in the case of the double well. }
  \label{fig:phase}
\end{figure}

 The results of the instanton analysis presented in this section are not valid for all values 
 of the parameter $\xi$ in the potentials (\ref{u4}) and (\ref{u3}). The reason is that, if $\xi$ is 
 sufficiently large, the classical instanton trajectory, given by the integral (\ref{eq:t-intx}) at $E=0$, 
 is no longer real, since the value of $u(x)$ in the integration interval is greater than $2$, and the 
 integral is complex.  It is easy to see that the critical values for $\xi$ occur at 
 \be 
 \label{crit-h}
 \xi_c=3, \qquad \xi_c=4, 
 \ee
 for the cubic potential and the double-well potential,
 respectively. For these values, $u(x)$ has a critical point (a local
 maximum) at which $u(x)=2$. Similarly, the action of the instanton
 (\ref{action-real}), which is real and positive for $\xi<\xi_c$, becomes
 complex.  An illustration of the shape of the double-well potential
 at $\xi<4$, $\xi=4$ and $\xi>4$ is shown in \figref{fig:phase}. As we will
 see, we can think of this as a {\it phase transition}, since the
 asymptotics as $\hbar \to 0$ of the non-perturbative effects changes
 discontinuously as $\xi$ varies through its critical value. The
 behavior at $\xi=\xi_c$ is also special: the scaling with $\hbar$ becomes
 ``anomalous", displaying a non-trivial power law, as we will see in
 detail in section \ref{sec:crit}. The approach to criticality can be
 also seen in the instanton trajectories shown in
 \figref{fig:dw-profiles} and \figref{fig:cw-profiles}. When we are
 very close to the critical point, they ``split" into smaller
 trajectories. The splitting point is the local maximum at which
 $u(x)=2$.
  
 We will call the phase $\xi<\xi_c$ the {\it strong coupling phase}, and
 the phase with $\xi>\xi_c$ the {\it weak coupling phase}. This is
 somewhat counter-intuitive but it is motivated by the following. As
 we noted in section \ref{sec:dQM}, the parameters in the potential
 are mapped to the moduli of the SW curve for the $SU(N)$ theory. It
 is easy to see that the values (\ref{crit-h}), together with the
 corresponding values of $\xi_{3,4}$ in \eqref{h3}, \eqref{h4} with
 $E=0$, correspond precisely to one of the {\it monopole points}
 \cite{sw,ds} in the moduli space of the $SU(3)$, $SU(4)$ SW curve,
 respectively (the one with real coordinates). These points mark the
 boundary between the strong coupling and the weak coupling phase of
 SW theory. There
 is actually a discontinuous change in SW theory at this point,
 namely, a jump in the BPS spectrum (or wall-crossing). As we will see
 in the next section, this jump is the ultimate
 responsible for the discontinuous asymptotics of the non-perturbative
 effects in deformed quantum mechanics.

  \subsection{From instantons to WKB}  
  \label{subsec-inst-wkb}
  
  The instanton results derived above are valid only for the ground state energy, since we considered 
  trajectories of classical vanishing energy. In order to obtain results for excited states it is more convenient to 
  use the exact WKB method (see \cite{iwaki} for a recent review). In the case of difference equations, 
  this method has not been developed in enough detail to allow a computation of the EQC, 
  but we can use the instanton results (\ref{egap-ground}), (\ref{ime-ground}) to obtain an educated guess for all 
  energy levels. Let us first review the situation for the conventional double and cubic wells, following \cite{aqm}. 
  
  Let us denote by 
  \be
  \Pi_{\rm WKB}(E; \hbar)= \sum_{\ell \ge 0} \Pi^{(\ell)}_{\rm WKB}(E) \hbar^{2 \ell}, 
  \ee
  the tunneling quantum period appearing in both the cubic and the
  double-well potentials. Its leading term is\footnote{\jjj{Added} Quantum periods are in general 
  complex, but in the cubic and the double-well models with $E$ real and slightly higher than
    the minimum of $V(x)$, the 1-cycles defining the periods reduce to intervals on the real line, so they are either real or purely imaginary. In the convention we are using, quantum periods corresponding to tunneling cycles are real.}
  \be
  \label{lead}
   \Pi^{(0)}_{\rm WKB}(E)=2 \int_{x_0(E)}^{x_\star^-(E)} {\sqrt{2  (V(x)-E)}} \rd x. 
   \ee
  Let us suppose that this is the only non-perturbative period that has to be considered in our problem 
  (as it happens in fact in conventional quantum mechanics). Then, at leading non-perturbative order, we have the 
  following formula for the 
  energy splitting of the $\nu$-th energy level in the double-well potential \cite{alvarez}: 
  \be
  \label{gap-wkb}
  \Delta E(\nu)\approx {1\over \pi} {\partial E \over \partial \nu} \re^{- \Pi_{\rm WKB}(E; \hbar)/\hbar}.  
  \ee
 Here, in the r.h.s., $E=E(\nu;\hbar )$ is the perturbative quantum energy as a function of the shifted quantum number 
  \be
  \label{nu-def}
  \nu= n+{1\over 2},  
  \ee
  and we have (see e.g. \cite{aqm})
  \be
  \label{Ehbar}
  E= \hbar \omega \nu + \CO(\hbar^2). 
  \ee
  An important subtlety is that, after using (\ref{Ehbar}), the higher orders in $\hbar$ of 
  $\Pi_{\rm WKB}/\hbar$ contain contributions which are $\hbar$-independent and singular at $\nu=0$\footnote{This has led to many confusions in the conventional WKB literature and to some erroneous formulae in famous textbooks. In particular, the formula for the energy gap of the double-well potential in \cite{landau} is not correct. An interesting discussion of these issues can be found in \cite{garg-tunnel}.}. More precisely, we can write 
  \be
  \Pi_{\rm WKB} (E(\nu, \hbar), \hbar)= \Pi^{\rm reg}_{\rm WKB} (E(\nu, \hbar), \hbar)+ \Pi^{\rm sing}_{\rm WKB} (E(\nu, \hbar), \hbar),
  \ee
  where 
  \be
  \label{nu-series}
  {1\over \hbar} \Pi^{\rm sing}_{\rm WKB} (E(\nu, \hbar), \hbar)=\nu (\log \nu-1)-{1\over 24 \nu} + {7 \over 2880 \nu^3} + \CO(\nu^5).  
  \ee
  The logarithmic term comes from the logarithmic singularity in (\ref{lead}), while the terms in $1/\nu^{2k-1}$ come from 
  $\Pi^{(k)}_{\rm WKB} (E)$ with $k\ge 1$. The correct prefactor in (\ref{gap-wkb}) 
  can be obtained by promoting the series above, after exponentiation,
  to 
  \be
  \label{G-pref}
  \re^{-\Pi^{\rm sing}_{\rm WKB}/\hbar} \rightarrow {\sqrt{2 \pi } \over \Gamma\left(\nu+{1\over 2}\right)}.   
  \ee
 This can be justified by an explicit all-orders calculation of the singular part of the period in the 
 exact WKB method \cite{ddpham,gaiotto}, 
 or by using the uniform WKB method \cite{alvarez} (see \cite{aqm} for a pedagogical 
 discussion.) Similarly, in the case of the cubic well, the imaginary 
 part of the energy for an arbitrary level $\nu$ is \cite{alvarez-casares2}
 \be
 \label{ime-wkb-qm}
 {\rm Im}\,  E(\nu)\approx -{1\over 4\pi} {\partial E \over \partial \nu} \re^{- \Pi_{\rm WKB}(E; \hbar)/\hbar}.
 \ee
 It conventional quantum mechanics, and when $\hbar \to 0$, the exact WKB formulae above lead to the usual instanton results. 
The formulae \eqref{egap-ground} and \eqref{ime-ground} are identical to the standard ones, except that they involve 
 the periods appropriate for deformed quantum mechanics. This suggests that the expressions (\ref{gap-wkb}) and (\ref{ime-wkb-qm}) are also valid in the WKB approach to deformed quantum mechanics, provided we use the quantum periods adapted to the finite difference equation (\ref{fde}). This means in particular that the leading non-perturbative contribution is only due to 
 the tunneling period. 
  
 Let us then briefly address the construction of formal quantum periods in deformed 
 quantum mechanics. This problem was already considered in \cite{dingle} and was 
 revisited more recently in \cite{delmonte, baldino}. In analogy with conventional quantum mechanics, 
 one uses the difference equation (\ref{fde}) to define a quantum 
 Liouville differential $p(x; \hbar) \rd x $, where $p(x; \hbar)$ is given by a formal power series
 \be
 p(x; \hbar) = \sum_{n \ge 0} p_n(x) \hbar^{2n}, 
 \ee
and 
 \be
 p_0(x)= \cosh^{-1}(Q(x)), \qquad Q(x)= u(x)-1 -E. 
 \ee
The higher order corrections can be computed systematically. For example, 
 \be
 \label{p1}
 p_1(x)=  - {1\over 12} {(Q^2(x)+ 2) Q''(x) \over (Q^2(x)-1)^{3/2} }+{Q(x) (Q'(x))^2 (13+2 Q^2(x)) \over 24 (Q^2(x)-1)^{5/2}}. 
 \ee
 Quantum periods are obtained by integration of this differential
 around the appropriate cycles. Tunneling periods are obtained by
 analytic continuation under the barriers, as in the usual WKB
 method. For example, the leading term of the tunneling period
 corresponding to (\ref{lead}) is
 \be
 \Pi(E)= 2 \int_{x_0(E)}^{x^-_\star(E)} \cos^{-1} \left( 1+ E-u(z) \right) \rd z.
 \ee
 Since we will not consider quantum corrections in this section, we will suppress the superscript in $\Pi$ that we used in (\ref{lead}). 
 
\begin{figure}
  \centering
  \subfloat[$\nu=1/2$]{\includegraphics[height=3cm]{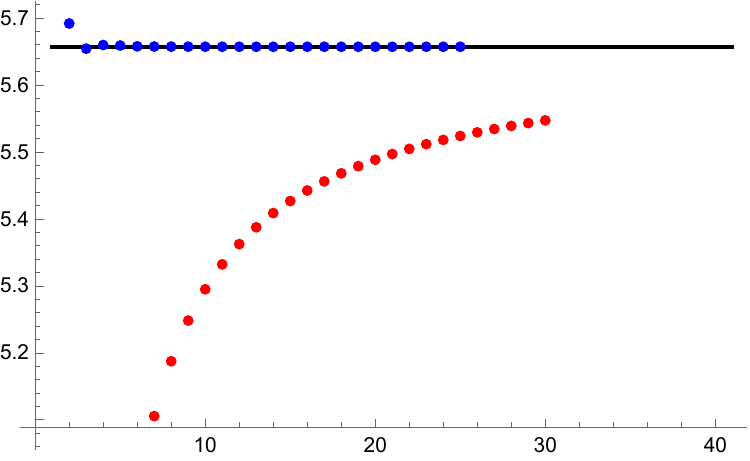}}%
  \hspace{18ex}
  \subfloat[$\nu=3/2$]{\includegraphics[height=3cm]{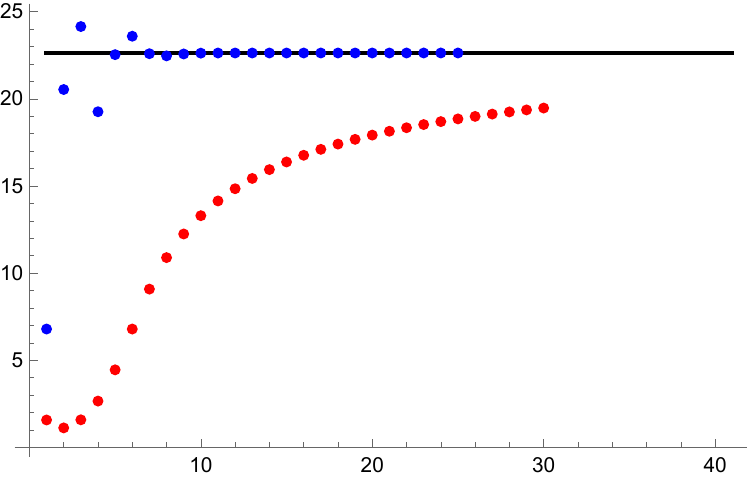}}%
  \caption{The energy difference $\Delta E$ at the first ($\nu=1/2$,
    left) and the second ($\nu=3/2$, right) energy levels of the
    double-well model with $\xi=2$ and $\hbar = 1/n$ ($n=1,2,\ldots$).
    The red dots are the numerical values of the energy differences
    divided by the exponential $\re^{-\CS/\hbar+2\nu \CA}$.  The blue
    dots are the results of Richardson transform to accelerate the
    numerical convergence.  The black lines are the analytic values
    from the prefactors in \eqref{eq:DelEnu-dw}.}
  \label{fig:DelE}
\end{figure}

\begin{figure}
  \centering
  \subfloat[$\nu=1/2$]{\includegraphics[height=3cm]{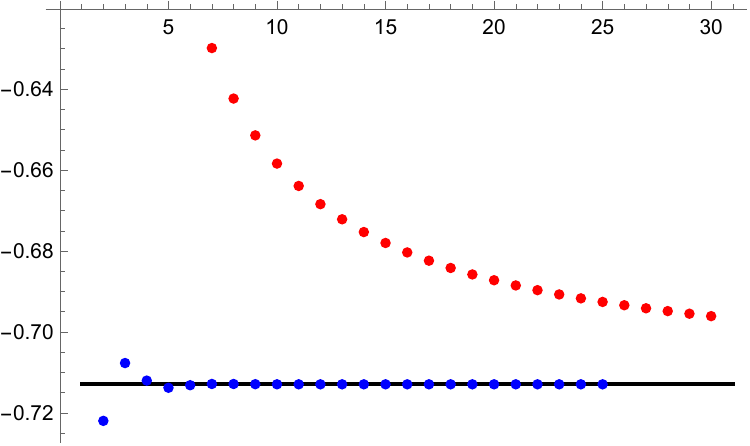}}\hspace{18ex}
  \subfloat[$\nu=3/2$]{\includegraphics[height=3cm]{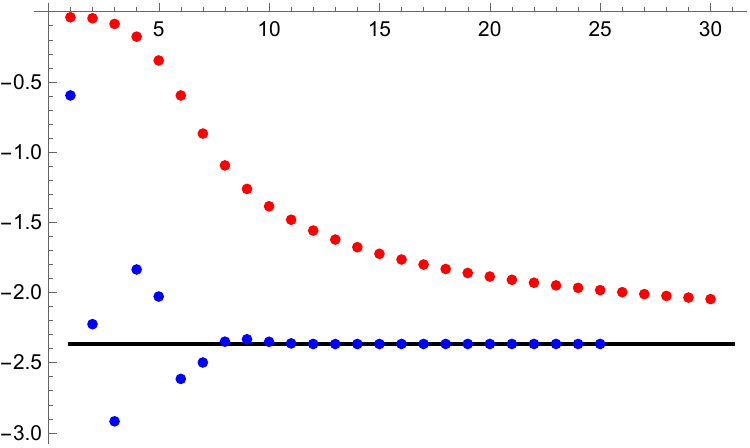}}
  \caption{Imaginary parts of the ground state energy ($\nu=1/2$,
    left) and the first excited state energy ($\nu=3/2$, right) of the
    cubic model with $\xi=1/2$ and $\hbar = 1/(2n)$ ($n=1,2,\ldots$).
    The red dots are the numerical values of the imaginary parts
    divided by the exponential $\re^{-\CS/\hbar+2\nu \CA}$.  The blue
    dots are the results of Richardson transform to accelerate the
    numerical convergence.  The black lines are the analytic values
    from the prefactors in \eqref{eq:ImEnu-cubic}.}
  \label{fig:ImE}
\end{figure}

Let us now compute the energy gap (\ref{gap-wkb}) and the imaginary part of the energy (\ref{ime-wkb-qm}) by 
using the periods of deformed quantum mechanics, instead of the conventional ones, in the limit 
$\hbar \rightarrow 0$ for fixed $\nu$. To do this we only need the classical period at order $E$ and the 
analogue of the quantum corrections (\ref{nu-series}). These turn out to be the same in the case of 
deformed quantum periods, since (\ref{nu-series}) captures the local information near the minima 
of the potential. This can be explicitly checked by using e.g. (\ref{p1}). Since 
$E$ is of order $\hbar$, we have to analyze the behavior of 
$\Pi(E)$ when $E=0$. As in the calculation of the period (\ref{eq:betaE}), this 
contains a logarithmic singularity 
in $E$ due to the coalescence of the two points $x_\star^\pm (E)$. To extract 
this singularity, as well as the leading term in the regular part, it is 
better to take first a derivative w.r.t. $E$, and then use the tricks in (\ref{split}) 
and (\ref{loge}). In this way one finds, 
\mmm{corrected!}
\be
\label{PiE0}
\Pi(E) =\CS+ {E\over \omega} \left( \log (2 E)-1 \right)-{2E \over \omega} \left( \CA + \log(2 \omega (x_\star -x_0)) \right) + \cdots,
\ee
where $\CA$ is given in (\ref{calA}). A simple calculation, after
taking into account that $\exp(-\nu(\log\nu-1))$ has to be promoted to
(\ref{G-pref}), gives
\be 
\label{eq:DelEnu-dw}
\Delta E (\nu)\approx -{\hbar \omega \over \pi} { {\sqrt{2 \pi}} \over \Gamma \left( \nu +{1\over 2} \right)} \left( {2 \omega (x_\star -x_0)^2 
\over \hbar} \right)^{\nu} \re^{-\CS/\hbar + 2 \nu \CA}
\ee
for the $\hbar \to 0$ asymptotics of the energy gap at the shifted level $\nu$
in the double well. For the imaginary part of the energy in the cubic potential we
find
\be
\label{eq:ImEnu-cubic}
 {\rm Im} \, E (\nu)\approx -{\hbar \omega \over 4 \pi} { {\sqrt{2 \pi}} \over \Gamma \left( \nu +{1\over 2} \right)} \left( {2 \omega (x_\star -x_0)^2 
\over \hbar} \right)^{\nu} \re^{-\CS/\hbar + 2 \nu \CA}. 
\ee
Both \eqref{eq:DelEnu-dw} and \eqref{eq:ImEnu-cubic} can be checked
against the numerical results by Hamiltonian matrix diagonalization,
and they are found to give the correct results with high precision, as
shown in Figs.~\ref{fig:DelE} and Figs.~\ref{fig:ImE}
respectively. \footnote{\jjj{Added} In producing these figures we have
  used the Richardson transform to improve the numerical convergence,
  and a standard reference can be found in
  \cite[Sec.~8]{bender-orszag}.}

\section{Weak coupling phase and suppression of tunneling}
\label{sec:weak}

\subsection{Complexifying the instantons}
\label{sec:comp}

As we mentioned in the previous section, the standard instanton analysis is only valid when the parameter $\xi$ in the potentials is smaller than a certain critical value. What happens when $\xi>\xi_c$? We can simply try to analytically continue the quantities characterizing the instanton past this value. We first note that the potential crosses the value $u(x)=2$ at two points in the intermediate region of the potential. 
We denote them by $x_{\rm np}^\pm (E)$, where ${\rm np}$ stands for ``non-perturbative" (the reasons for this name is that these points characterize the new tunneling periods, as we will see). These points satisfy 
\be
u\left( x_{\rm np}^\pm (E) \right)= 2.
\ee
 In the double well they are given by 
 \be
 x^\pm_{\rm np}(E) = \pm{\sqrt{{\xi \over 2}- {\sqrt{4+ E}}}},
 \ee
 and in the cubic potential they satisfy
\be
x_0(E) < x_{\rm np}^-(E)<x_{\rm np}^+(E) <x_\star(E)
\ee
(we recall that in this potential $x_0(E)$ is the turning point for the energy $E$). Like before, 
we will denote by $x_{\rm np}^\pm = x_{\rm np}^\pm (0)$ the values at zero energy. The action (\ref{action-real}) now becomes complex, since the integrand is complex in the interval $[x^-_{\rm np}, x^+_{\rm np}]$, and we find 
\be
\CS= \CS_{\IR} + \ri \CS_{\CI}, 
\ee
where 
\be
\CS_\IR=  2 \int_{x_{\rm np}^+}^{x_\star} \cos^{-1} \left(1-u(x) \right) \rd x +  2 \pi x_{\rm np}^+, \qquad S_\CI=-2 \int_0^{x_{\rm np}^+} \cosh^{-1} \left(u(x)-1 \right) \rd x, 
\ee
for the double well, and 
\be
\ba
\CS_\IR&=  2 \left( \int_{x_0}^{x_{\rm np}^-} +\int_{x_{\rm np}^+}^{x_\star} \right) \cos^{-1} \left(1-u(x) \right) \rd x +  2 \pi \left(x_{\rm np}^+- x_{\rm np}^- \right),\\
 \CS_\CI&=-2 \int_{x_{\rm np}^-}^{x_{\rm np}^+} \cosh^{-1} \left(u(x)-1 \right) \rd x,
\ea
\ee
for the cubic potential. A similar complexification occurs for the quantity $\CA$ in (\ref{calA}) appearing in the 
one-loop correction to the instanton amplitude, and we will write it as 
\be
\CA = \CA_{\IR} + \ri \CA_{\CI}. 
\ee
We have 
\be
\label{aRI}
\CA_{\IR} =  \int_{x_{\rm np}^+}^{x_\star}\frac{\omega\rd x}{\sqrt{u(x)(2-u(x))}} -
  \int_{x_0}^{x_\star}\frac{\rd x}{x_\star-x}, \qquad \CA_\CI=   \int_0^{x_{\rm np}^+}\frac{\omega\rd x}{\sqrt{u(x)(u(x)-2)}}, 
  \ee
  for the double well, and 
 \be
 \ba
\CA_{\IR} &=  \left( \int_{x_0}^{x_{\rm np}^-} +\int_{x_{\rm np}^+}^{x_\star} \right) \frac{\omega\rd x}{\sqrt{u(x)(2-u(x))}} -
  \int_{x_0}^{x_\star}\frac{\rd x}{x_\star-x},\\
   \CA_\CI&=  \int_{x_{\rm np}^-}^{x_{\rm np}^+} \frac{\omega\rd x}{\sqrt{u(x)(u(x)-2)}}, 
   \ea
  \ee
for the cubic well\footnote{The sign of $\CA_\CI$ involves a choice of sign in the square root in (\ref{aRI}). This can be fixed by 
noting that $\CA_I$ is the subleading term in the expansion of the energy-dependent WKB period around $E=0$.}.

Since the instantons have become complex, they have an imaginary phase
which might lead to suppression of tunneling, as it happens e.g. in
the spin systems studied in \cite{loss, garg}. We note that in
conventional quantum mechanics, and for the potentials we are
considering, complex instantons do not occur for the low energy
levels, and tunneling is never suppressed\footnote{Complex instantons
  appear and play a r\^ole in the double-well for energies above the
  barrier, see e.g. section 5.2 of \cite{ims}, but this is a very
  different situation from what we have here, since classical motion
  between the critical points is no longer forbidden in that
  case.}. The existence of complex instantons in deformed quantum
mechanics is ultimately due to the unconventional kinetic term in the
Hamiltonian. The complexification of the instantons can also be seen
in the Euclidean EOM (\ref{eq:dyn}) and Figs.~\ref{fig:w4dec},
\ref{fig:w3dec}, involving the effective inverted potential $w(x;E)$:
as $\xi$ grows to its critical value, the bump in the middle of the
well for $w(x;E)$ grows up and ``cuts" the instanton trajectory.

\begin{figure}
  \centering
  \subfloat[$\nu=1/2$]{\includegraphics[height=3.5cm]{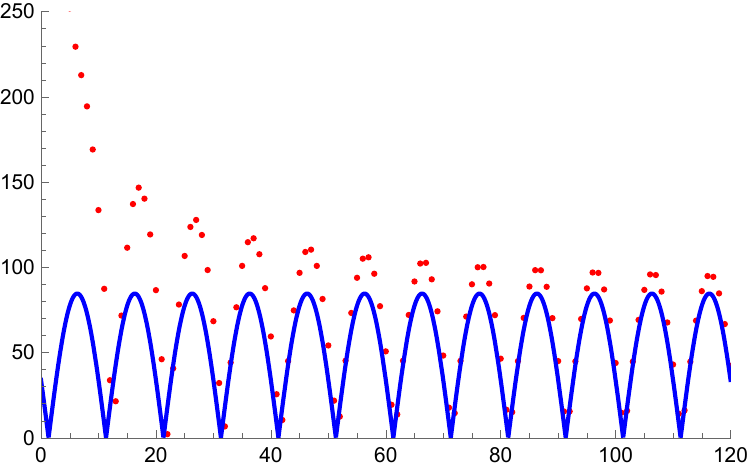}%
    \label{fig:dw-E0-oscT1}}%
  \hspace{12ex}
  \subfloat[$\nu=3/2$]{\includegraphics[height=3.5cm]{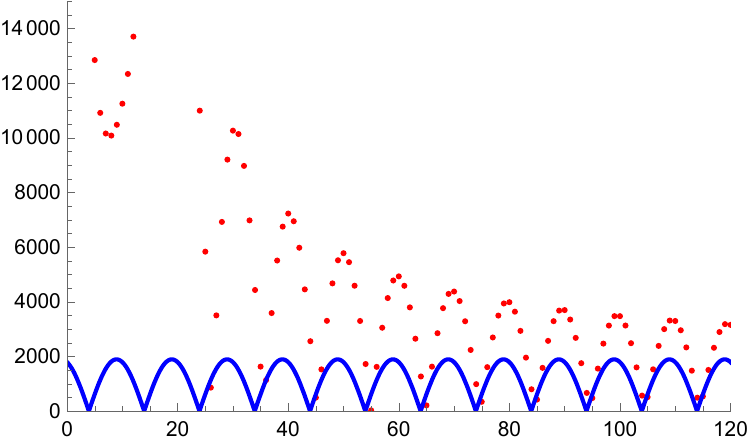}%
    \label{fig:dw-E0-oscT2}}%
  \caption{The function $\Delta\mc{E}(\nu,\hbar)$ evaluated on the
    sequence $\hbar^{(1)}_n$ for (a) $\nu=1/2$ and (b) $\nu=3/2$ with
    $\xi = 10$.  The red dots are numerical results, and the blue curve
    is the prediction from complexification of instantons.}
  \label{fig:dw-oscT1} 
\end{figure}

\begin{figure}
  \centering
  \subfloat[$\nu=1/2$]{\includegraphics[height=3cm]{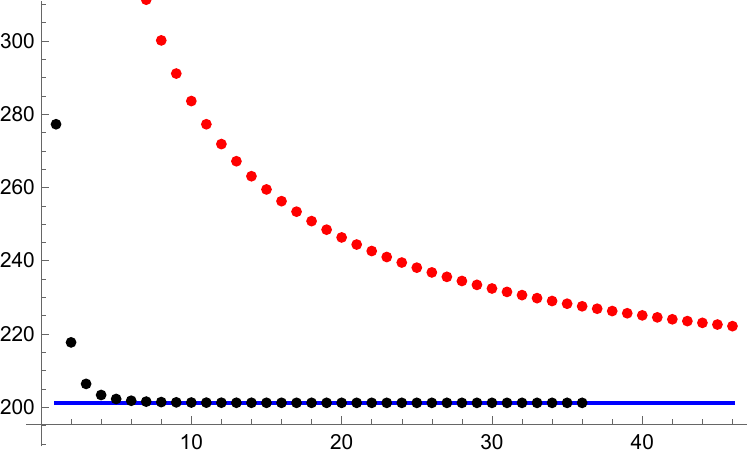}%
    \label{fig:dw-E1-oscT1}}%
  \hspace{14ex} 
  \subfloat[$\nu=3/2$]{\includegraphics[height=3cm]{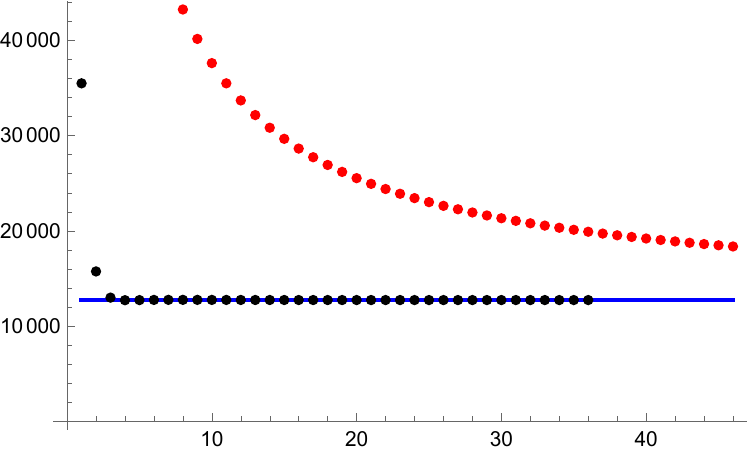}%
    \label{fig:dw-E1-oscT2}}%
  \caption{The function $\Delta\mc{E}(\nu,\hbar)$ evaluated on the
    sequence $\hbar^{(2)}_n$ for (a) $\nu=1/2$ and (b) $\nu=3/2$ with
    $\xi = 20$.  The red dots are numerical results, the black dots are
    results after applying the Richardson transform to speed up the
    convergence, and the blue line is the asymptotic value predicted
    from complexification of instantons.}
  \label{fig:dw-oscT2}
\end{figure}

Let us now see how we can incorporate these complex instantons in the
calculation of the non-perturbative effects. One obvious approach is to
add up the contribution of the complex instanton and its complex
conjugate. This is the procedure suggested in the seminal paper
\cite{blgzj} and reviewed in section 3.5 of \cite{mmbook}. When
applied to the double-well potential, we find the result
\be
\label{eq:complex-dw}
\Delta E (\nu)\approx -{2\hbar \omega \over \pi} { {\sqrt{2 \pi}} \over \Gamma \left( \nu +{1\over 2} \right)} \left( {2 \omega (x_\star -x_0)^2 
\over \hbar} \right)^{\nu} \cos\left( {\CS_\CI \over \hbar} - 2 \nu \CA_\CI \right) \re^{-\CS_\IR/\hbar + 2 \nu \CA_\IR}.
\ee
This formula leads indeed to suppression of tunneling, due to the $\cos$
term. We note that, although we have not indicated it explicitly, 
the energy gap is given, strictly speaking, by the absolute value of the 
above expression, since when suppression of tunneling happens, the roles 
of the ground state and the first excited state are interchanged. 
We will verify in a moment that this is indeed the instanton
approximation to the exact formula conjectured in \cite{gm-dqm}. In addition, a direct calculation of the spectrum of the Hamiltonian
for $\xi>\xi_c$ shows that this formula describes the correct asymptotic
behaviour of the energy gap at small $\hbar$.  Let us evaluate
numerically the function
\begin{equation}
  \Delta\mc{E}(\nu,\hbar) :=
  |(2/\hbar)^{1-\nu}(\pi/2)^{1/2}\Gamma(\nu+1/2)
  \re^{\CS_\IR/\hbar-2\nu \CA_\IR}
  \Delta  E(\nu,\hbar)|
\end{equation}
on a sequence of $\hbar$,
\begin{equation}
  \hbar^{(1)}_n= -10 \CS_\CI/(\pi n ),\quad n = 1,2,3,\ldots.
\end{equation}
Eq.~\eqref{eq:complex-dw} predicts that in the large $n$ limit the value of
$\Delta\mc{E}(\nu,\hbar^{(1)}_n)$ should oscillate following the
pattern
\begin{equation}\label{eq:DelE-oscT1}
  \Delta\mc{E}(\nu,\hbar^{(1)}_n)\approx
  4\omega^{1+\nu}(x_\star-x_0)^{2\nu}|\cos\left(-(\pi/10) n-2\nu\CA_\CI\right)|.
\end{equation}
As illustrated in Figs.~\ref{fig:dw-oscT1}, we find that this is
indeed the case for both $\nu=1/2$ and $\nu=3/2$.  The numerical value
of $\Delta\mc{E}(\nu,\hbar^{(1)}_n)$ and the right hand side of
\eqref{eq:DelE-oscT1} oscillate locked in phase and they reach crests
and troughs at the same time.  In addition, the amplitude of
oscillation of $\Delta\mc{E}(\nu,\hbar^{(1)}_n)$ seems to asymptote to
that of the right hand side of \eqref{eq:DelE-oscT1} in the large $n$
limit.  To make a higher precision check of the latter statement, we
evaluate $\Delta\mc{E}(\nu,\hbar)$ on a second sequence of $\hbar$ \jjj{Corrected!}
\begin{equation}
  \hbar^{(2)}_n = - \CS_\CI/(\pi n + 2\nu\CA_\CI), \quad n=1,2,3,\ldots.
\end{equation}
According to~\eqref{eq:complex-dw} its value should asymptote to the maximum value of the
right hand side of \eqref{eq:DelE-oscT1},
\begin{equation}
  \Delta\mc{E}(\nu,\hbar^{(2)}_n) \approx 4\omega^{1+\nu}(x_\star-x_0)^{2\nu},
\end{equation}
which is indeed verified, as shown in Figs.~\ref{fig:dw-oscT2}.

However, if we apply the same procedure to the cubic potential, and we
simply take into account the contribution of the complex instanton and
its conjugate, the resulting formula for the imaginary part of the
energy turns out to be incorrect. This means that the structure of the
instantons of the theory is more complicated. In particular, naive
complexification fails. We will now explain why this is so and why the
naif approach is not guaranteed to be valid in general.

\subsection{Resurgent analysis and Stokes constants}

Determining the correct instanton content in a quantum mechanical
model can be challenging. The conventional picture of instantons as
classical solutions of the Euclidean EOM, which we used successfully
in section \ref{sec:strong}, does not always give a reliable guide, in
particular in situations where complex instantons arise. We have just
seen that the naif complexification of the instantons in the cubic
potential seems to miss crucial information. One needs a more general
setting which, in conventional one-dimensional quantum mechanics, was
put forward in the seminal work of Voros and collaborators, also
called the exact complex WKB method \cite{bpv,voros-quartic}. The first lesson of this method is that, in order to
obtain the collection of possible non-perturbative corrections or
trans-series, one should consider a lattice of instanton actions in
the complex plane (or more generally, a lattice of complex periods of
the underlying classical Hamiltonian, regarded as an algebraic
curve). To each point $\CS$ of this lattice one can associate
trans-series weighted by the exponentially small term
$\exp(-\CS/\hbar)$. In order to determine which are the instantons in
this lattice that contribute to a given quantity, one has to
perform a detailed analysis which involves the Voros--Silverstone
connection formula for the Schr\"odinger operator
\cite{voros-quartic,silverstone} (see \cite{aqm} for a pedagogical
exposition).

In the context of difference equations, the precise analogue of the
Voros--Silverstone connection formula is not known, although it is
expected to be more complicated.  To have a handle on our problem we
will use a shortcut and appeal to the theory of resurgence. According
to this theory, some of the non-perturbative corrections which are
needed in a problem (but not necessarily all) can be read from the large order behaviour of
perturbation theory (see \cite{mmbook} for a readable introduction to
the subject). This method can fail in special situations, mostly due to the presence of 
symmetries. For example,
in the double-well potential in conventional quantum mechanics, the large order behavior 
of the perturbative series misses the one-instanton correction to the energy gap 
(as we will see, this is also the case in the deformed double-well). However, 
in the case of the cubic well potential, the instanton effect responsible for the imaginary part of the energy is completely
captured by the large order behavior of the perturbative series, or
equivalently, by the structure of Borel singularities of its Borel
transform. We will then analyze the resurgent structure of the
perturbative series in deformed quantum mechanics to extract the
missing information on the instanton contributions.

We first give a lightning review of the basic ideas in the theory of resurgence, mostly to fix the notation. 
Let us consider a formal series
\begin{equation}
  \phi(z) = \sum_{n=0}^\infty a_n z^n,\quad a_n \approx n!,
\end{equation}
whose coefficients grow factorially. Its Borel transfom 
\begin{equation}
  \wh{\phi}(\zeta) = \sum_{n=0}^\infty\frac{a_n}{n!}\zeta^n,
\end{equation} 
has a finite radius of convergence and in favourable circumstances can
be analytically continued to the complex planes, where it will have
singularities. We also assume that
$\wh{\phi}(\zeta) \lesssim \exp(|\zeta|/R)$ i.e. it grows at most
exponentially fast \jjj{Modified} as $\zeta\rightarrow \infty$. The
Borel resummation of the original series $\phi(z)$ is then given by
\begin{equation}\label{eq:borel}
  \mr{S}\phi(z) = \frac{1}{z}\int_0^\infty
  \wh{\phi}(\zeta)\re^{-\zeta/z}\rd \zeta,
\end{equation}
The existence of the Borel
resummation depends crucially on the distribution of the singular
points of $\wh{\phi}(\zeta)$ in the complex $\zeta$-plane, called the
Borel plane. If $\wh{\phi}(\zeta)$ has singularities on the
positive real axis, \eqref{eq:borel} is ill-defined and
the original series $\phi(z)$ is said to be not Borel summable. Instead, one defines a pair of
lateral Borel resummations,
\begin{equation}\label{eq:lateral}
  \mr{S}^{(\pm)}\phi(z) = \frac{1}{z}\int_0^{\re^{\pm\ri 0^+}\infty}
  \wh{\phi}(\zeta)\re^{-\zeta/z}\rd \zeta.
\end{equation}
This definition can be generalized to
\begin{equation}\label{eq:theta}
  \mr{S}^{(\theta)}\phi(z) = \frac{1}{z}\int_0^{\re^{\ri \theta}\infty}
  \wh{\phi}(\zeta)\re^{-\zeta/z}\rd \zeta.
\end{equation}
for $|\theta| < \pi/2$, and it coincides with the positive (negative)
lateral resummation when $\theta >0$ (resp.~$\theta <0$) if the Borel
transform $\wh{\phi}(\zeta)$ has no additional singularities between
the positive real axis $\IR_+$ and the ray
$\rho_\theta = \re^{\ri\theta}\IR_+$ in the Borel plane.  Due to the
singular points on the positive real axis, the two lateral
resummations differ by a non-perturbatively small quantity, and their
difference, known as the Stokes discontinuity\footnote{The subscript 0
  refers to the branch of zero inclination angle across which the
  discontinuity is calculated.}
\begin{equation}
  \text{disc}_0\phi(z) = \mr{S}^{(+)}\phi(z) - \mr{S}^{(-)}\phi(z),
\end{equation}
is controlled by these singular points.  Let the singularities be
located at
$\CS_1 < \CS_2 < \ldots$, and let us suppose that each of them is associated with an
additional power series
\begin{equation}
  \varphi^{(i)}(z) = \sum_{n=0}^{\infty} a_n^{(i)}z^n. 
\end{equation}
Then, the Stokes discontinuity is given by
\begin{equation}\label{eq:disc-phi}
  \text{disc}_0\phi(z) = \sum_{i\geq 1} \mS_i
  \mr{S}^{(-)}\phi^{(i)}(z),\quad
  \phi^{(i)}(z)=z^{-\nu_i}\re^{-\CS_i/z}\varphi^{(i)}(z),
\end{equation}
where the Borel resummation on the right hand is applied on the power
series $\varphi^{(i)}(z)$. The coefficients $\mS_i$ are known as
Stokes constants, and their value depends on a choice of normalization for 
$\varphi^{(i)}$.  Clearly, the
distribution of singular points of $\wh{\phi}(\zeta)$, the power
series associated to them, as well as the Stokes constants, are
crucial for making sense of the asymptotic series $\phi(z)$ via Borel
resummation, and they are known collectively as the resurgence
structure of $\phi(z)$.

\begin{figure}
  \centering
  \subfloat[$\nu=1/2,\xi=3/4$]{\includegraphics[height=4cm]{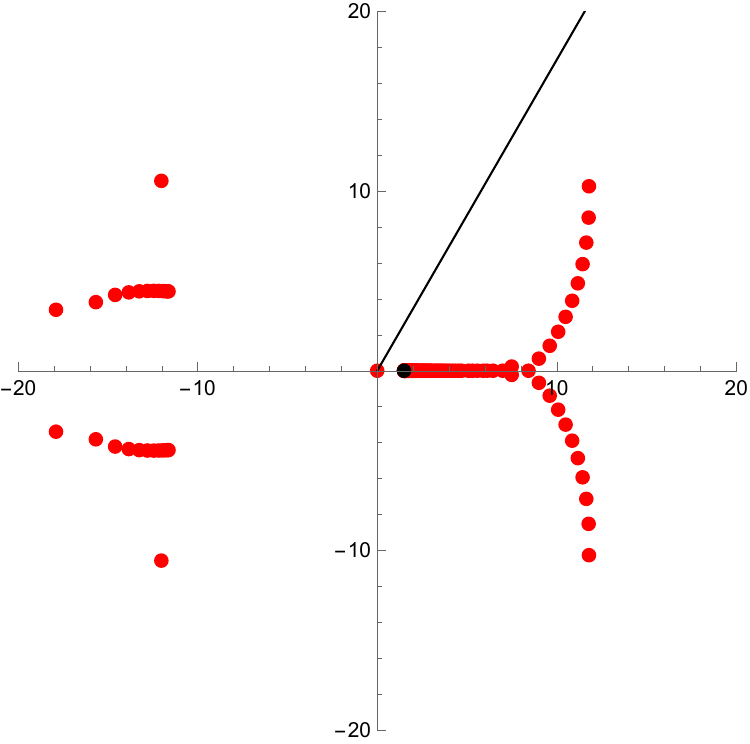} \label{fig:hsmall-brl}}\hspace{12ex}
  \subfloat[$\nu=1/2,\xi=12$]{\includegraphics[height=4cm]{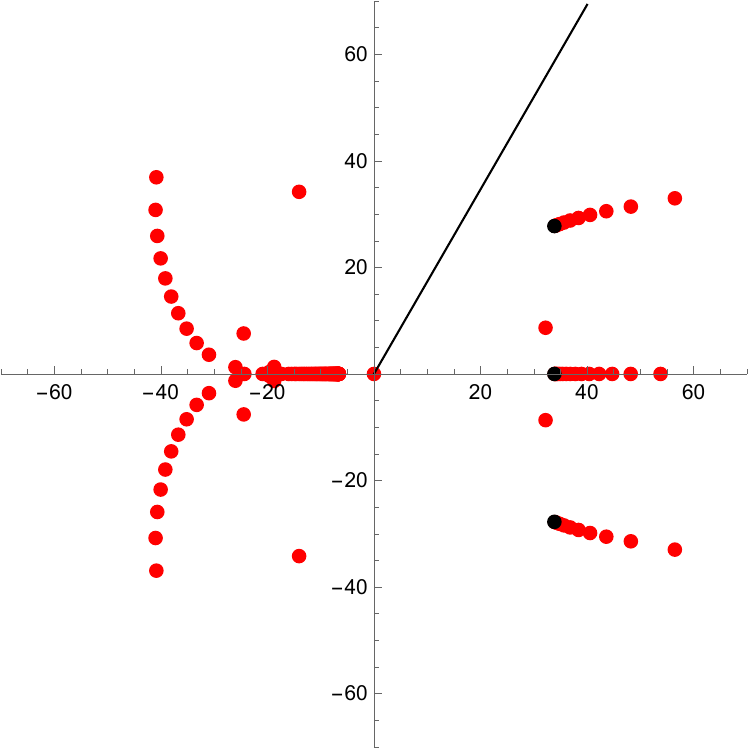} \label{fig:hlarge-brl}}\\
  \subfloat[$\nu=3/2,\xi=3/4$]{\includegraphics[height=4cm]{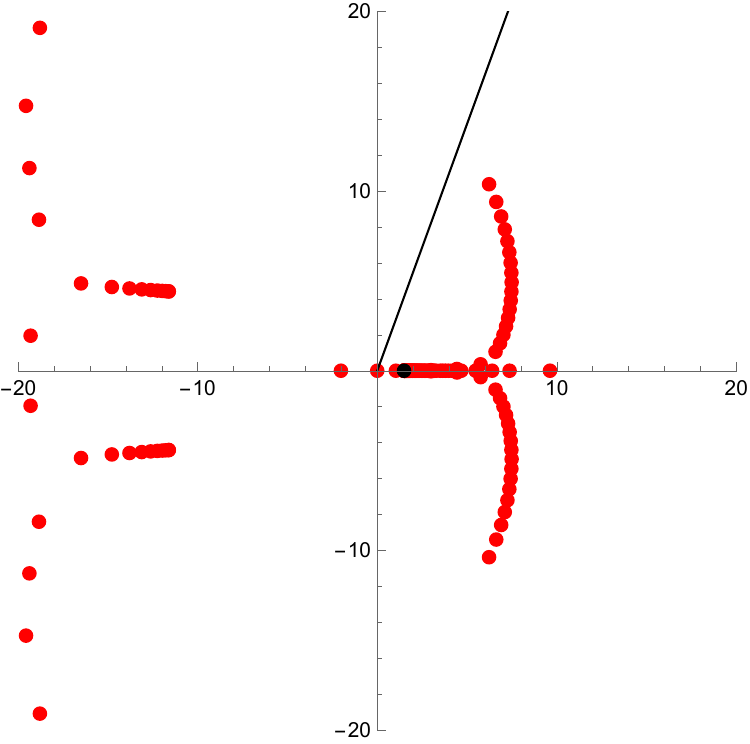} \label{fig:E1-hsmall-brl}}\hspace{12ex}
  \subfloat[$\nu=3/2,\xi=12$]{\includegraphics[height=4cm]{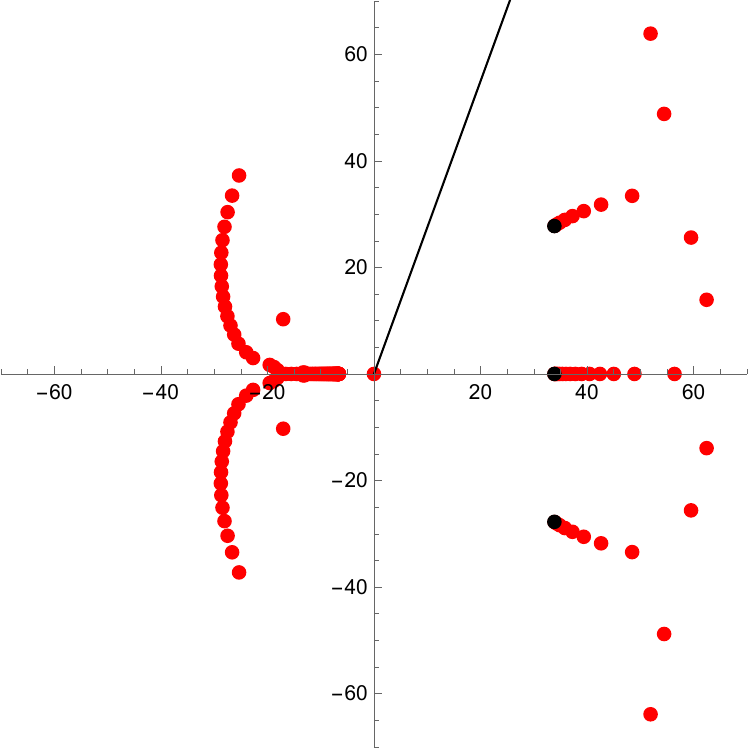} \label{fig:E1-hlarge-brl}}
  \caption{The Borel plane of the perturbative energy series of the
    cubic model at both $\nu=1/2$ (upper) and $\nu=3/2$ (lower) for
    both $\xi<\xi_c$ (left) and $\xi>\xi_c$ (right). The red dots are position
    of poles of Pad\'e approximant of the Borel transform of the
    perturbative series truncated up to 300 terms. The black dots are
    the position of the branch points. The black lines are the
    integration contour for the (upper) lateral Borel resummation,
    which are raised at $60^\circ$ for $\nu=1/2$ and at $70^\circ$ for
    $\nu=3/2$.}
  \label{fig:cubic-brl}
\end{figure}
%

Let us now explore the resurgent structure of our models. We first
look at the cubic model with Hamiltonian $\mH_3$ \eqref{eq:HN}. Let us
consider the perturbative series for the energy
$E^{\text{pert}}(\nu,\hbar)$ at fixed shifted level $\nu$. This series
can be efficiently calculated to hundreds of terms by using the
\texttt{Mathematica} package \texttt{BenderWu}
\cite{package,tim-jie}. We find
\begin{equation}
  E^{\text{pert}}(\nu,\hbar) = (3\xi)^{1/4}\nu \hbar
  +\left(\left(-\frac{7}{192\xi}
      +\frac{\sqrt{3\xi}}{64}\right)+\left(-\frac{5}{16\xi}
      +\frac{\sqrt{3\xi}}{16}\right)\nu^2\right)
  \hbar^2 + \CO(\hbar^3).
\end{equation}

We first look at the case with $\xi<\xi_c$.  The Borel planes for
$E^{\text{pert}}(\nu,\hbar)$ at $\nu=1/2$ and $\nu=3/2$ are shown in
Figs.~\ref{fig:hsmall-brl}, \ref{fig:E1-hsmall-brl}.  Singular points
are found on the positive real axis, so that the perturbative
series $E^{\text{pert}}(\nu,\hbar)$ is not Borel summable.  On the
other hand, as demonstrated in Tab.~\ref{tab:Eisum-hsmall}, we find
that the lateral Borel resummation
$\mr{S}^{(+)}E^{\text{pert}}(\nu,\hbar)$ agrees with the numerical
evaluation of the energy $E(\nu,\hbar)$ obtained by Hamiltonian
matrix diagonalization:
\begin{equation}
  \mr{S}^{(+)}E^{\text{pert}}(\nu,\hbar) =
  \mr{S}^{(\theta_+)}E^{\text{pert}}(\nu,\hbar) = E(\nu,\hbar),\quad \xi<\xi_c.
\end{equation}
This is similar to what happens in the cubic well in conventional quantum mechanics. 
Note that in the evaluation of
$\mr{S}^{(+)}E^{\text{pert}}(\nu,\hbar)$ we need to significantly
raise the inclination angle $\theta_+$ of the integration contour in
order to avoid all the singular points that trail off from those on
the positive real axis\footnote{Note that moving to the right in the
  Borel plane the trail of singular points splits to two uncanny
  branches, which stem from our limited precision in calculating the
  Borel transform $\wt{E}^{\text{pert}}(\nu,\zeta)$ by using a
  truncated perturbative series and Pad\'e approximant as means for
  analytic continuation.}, and this is emphasized by the notation
$\mr{S}^{(\theta_+)}E^{\text{pert}}(\nu,\hbar)$ in the middle of the
identity above.  From the definition of lateral resummation
\eqref{eq:lateral}, we also have
\begin{equation}\label{eq:S-}
  \mr{S}^{(-)}E^{\text{pert}}(\nu,\hbar) =
  \overline{E}(\nu,\hbar),\quad \xi<\xi_c,
\end{equation}
so that 
\begin{equation}
  \imag \,  E(\nu,\hbar) = \frac{1}{2\ri}\text{disc}_0 E^{\text{pert}}(\nu,\hbar)
  = \frac{1}{2\ri}\mr{S}^{(-)}E^{(1)}(\nu,\hbar) + \ldots,\quad \xi<\xi_c,
\end{equation}
and we have chosen a normalization so that $\mS_1 = 1$ for the leading singular point.  In
the small $\hbar$ limit, the imaginary part of the ground state energy
is related to the non-perturbative corrections by
\begin{equation}
  \imag \, E(\nu,\hbar) \approx \frac{1}{2\ri}
  E^{(1)}(\nu,\hbar),\quad \xi<\xi_c.
\end{equation}
By numerical evaluation, we find indeed that 
\begin{equation}
  E^{(1)}(\nu,\hbar)
  \approx -\frac{\ri\hbar \omega}{\sqrt{2\pi} \Gamma \left( \nu + \frac{1}{2}
    \right)}
  \left( \frac{2 \omega (x_\star -x_0)^2}{\hbar} \right)^{\nu} \re^{-\CS/\hbar + 2 \nu \CA}
  ,\quad \xi<\xi_c,
\end{equation}
in agreement with \eqref{eq:ImEnu-cubic}, which is reassuring.
Note that the right hand side of the formula above has the same form
as $\phi^{(i)}$ in \eqref{eq:disc-phi}, and the instanton action $\CS$
agrees with the location of the leading singular point in
Fig.~\ref{fig:hsmall-brl}, indicating that the latter plays the role
of the branch point at the head of a branch cut.
Furthermore, there are also singular points on the left half of the
Borel plane, but they are irrelevant from the definition of the
lateral resummations \eqref{eq:lateral} and \eqref{eq:theta} as long
as we are only concerned with the Borel resummation of the
perturbative energy with $\hbar > 0$.

\begin{table}
  \centering
  \begin{tabular}{*{3}{>{$}l<{$}}}\toprule
    \nu=1/2
    & \hbar = 1
    & \hbar = 1/3 \\\midrule
    \mr{S}^{(+)}E^{\text{pert}}(\hbar)
    & 0.53409747428 - \ri\,0.19689238266
    & 0.18177567215412 - \ri\,0.00914131410440\\
    E^{\text{num}}(\hbar)
    & 0.53409747425 - \ri\,0.19689238261
    & 0.18177567215419 - \ri\,0.00914131410448\\
    E^{(1)}(\hbar)
    & \CO(0.1)
    & \CO(0.01)\\\midrule
    \nu=3/2
    & \hbar = 1
    & \hbar = 1/3 \\\midrule
    \mr{S}^{(+)}E^{\text{pert}}(\hbar)
    & 1.804970 - \ri\, 1.168308
    & 0.505013430505 - \ri\,0.137929992127
    \\
    E^{\text{num}}(\hbar)
    & 1.804960 - \ri\, 1.168316
    & 0.505013430512 - \ri\,0.137929992119
    \\
    E^{(1)}(\hbar)
    & \CO(1)
    & \CO(0.1)\\    
    \bottomrule
  \end{tabular}
  \caption{Comparison of the (upper) lateral Borel resummed
    perturbative energy series
    $\mr{S}^{(+)}E^{\text{pert}}(\nu,\hbar)$ and the numerical
    evaluation of the eigen-energy $E^{\text{num}}(\nu,\hbar)$ at
    $\nu=1/2$ and $\nu=3/2$ for $\xi=3/4$. The lateral resummation is
    performed with perturbative series truncated to 300 terms and with
    integration contour inclined at $60^\circ$ ($\nu=1/2$) and
    $70^\circ$ ($\nu=3/2$).  The numerical evaluation is performed by
    diagonalizing the Hamiltonian matrix using the eigenstates of
    harmonic oscillator with $\omega=1$ as basis vectors truncated to
    rank 100. For reference, the estimates of 1-instanton correction
    $E^{(1)}(\hbar)$ calculated by \eqref{eq:E1} are also given.}
  \label{tab:Eisum-hsmall}
\end{table}

Next, we switch to the case when $\xi>\xi_c$.  The Borel plane is given in
Fig.~\ref{fig:hlarge-brl}.  In contrast to the previous case, the
single singular point on the positive real axis splits to {\it three singular point}s, where the
additional two are above and below the positive real axis.  It turns
out that the suitable way to define the (upper) lateral Borel
resummation is to raise the inclination $\theta_+$ of the integration
contour high enough so that it goes over all the three branch
cuts
, as indicated in
Fig.~\ref{fig:hlarge-brl}.  Then we find numerically
\begin{equation}
  \mr{S}^{(+)}E^{\text{pert}}(\nu,\hbar):=
  \mr{S}^{(\theta_+)}E^{\text{pert}}(\nu, \hbar)
  = E(\nu,\hbar),\quad \xi>\xi_c,
\end{equation}
as demonstrated in Tab.~\ref{tab:Eisum-hlarge}.

This is actually quite natural from the resurgence point of view.  We
can fix the inclination angle $\theta_+$ of the integral contour for
the (upper) lateral resummation so that the contour is both above the
branch cut along the positive real axis (as well as the spurious
poles) when $\xi<\xi_c$ and over all the three branch cuts when $\xi>\xi_c$,
as in Fig.~\ref{fig:cubic-brl}.  When we change continuously the value
of $\xi$ from smaller than $\xi_c$ to greater than $\xi_c$, no singular
points of the Borel transform $\wh{E}^{\text{pert}}(\nu,\zeta)$ cross
the integral contour in the Borel plane, and consequently the value of
the (upper) lateral resummation
$\mr{S}^{(\theta_+)}E^{\text{pert}}(\nu,\hbar)$ changes continuously.
As the exact value of $E(\nu,\hbar)$ also changes continuously in the
process, these two functions should agree both when $\xi<\xi_c$ and when
$\xi>\xi_c$.

Using the same trick as \eqref{eq:S-}, we then conclude that
\begin{align}
  \imag \, E(\nu,\hbar)
  &= \mr{S}^{(+)}E^{\text{pert}}(\nu,\hbar) -
    \mr{S}^{(-)}E^{\text{pert}}(\nu,\hbar)\nn\\
  &\approx \frac{1}{2\ri} \left(\mS_1
    E^{(1)}(\nu,\hbar) + \mS_{1+}E^{(1+)}(\nu,\hbar) +
    \mS_{1-}E^{(1-)}(\nu,\hbar) \right) + \ldots,\quad \xi>\xi_c,
\end{align}
where $E^{(1)}(\nu,\hbar), E^{(1+)}(\nu,\hbar), E^{(1-)}(\nu,\hbar)$ are
associated to the singularities on, above, and below the positive real
axis respectively.
By evaluating numerically the Stokes discontinuities across these
three individual branch cuts, we find that
\begin{subequations}\label{eq:E1}
  \begin{align}
    &E^{(1)}(\nu,\hbar) =
      -\frac{\ri\hbar \omega}{\sqrt{2\pi} \Gamma \left( \nu + \frac{1}{2}
      \right)}
      \left( \frac{2 \omega (x_\star -x_0)^2}{\hbar} \right)^{\nu}\re^{-\CS_\IR/\hbar+2\nu\CA_\IR},\\
    &E^{(1\pm)}(\nu,\hbar) =
      -\frac{\ri\hbar \omega}{\sqrt{2\pi} \Gamma \left( \nu + \frac{1}{2}
      \right)}
      \left( \frac{2 \omega (x_\star -x_0)^2}{\hbar} \right)^{\nu}\re^{-(\CS_\IR\pm \ri \CS_\CI)/\hbar+2\nu(\CA_\IR\pm\ri\CA_\CI)},
  \end{align}
\end{subequations}
and the associated Stokes constants are 
\begin{equation}
  \mS_1 =  2,\quad \mS_{1\pm} = 1.
\end{equation}
Consequently, we should have
\begin{equation}
  \label{imr-ground}
  \imag\,E(\nu,\hbar) \approx
  -\frac{\hbar \omega}{\sqrt{2\pi} \Gamma \left( \nu + \frac{1}{2}
    \right)}
  \left( \frac{2 \omega (x_\star -x_0)^2}{\hbar} \right)^{\nu}
  \left(1+\cos\left({\CS_\CI \over \hbar}-2\nu\CA_\CI\right)\right)\re^{-\CS_\IR/\hbar+2\nu\CA_\IR}.
\end{equation}

\begin{figure}
  \centering
  \subfloat[$\nu=1/2$]{\includegraphics[height=3.5cm]{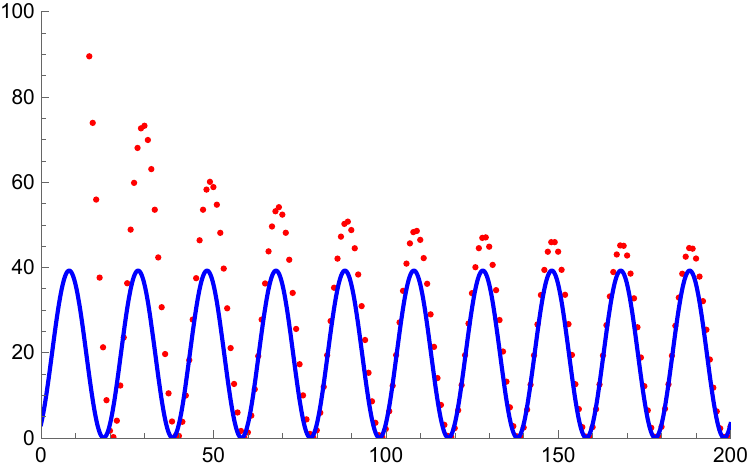}%
    \label{fig:cubic-E0-oscT1}}%
  \hspace{12ex}
  \subfloat[$\nu=3/2$]{\includegraphics[height=3.5cm]{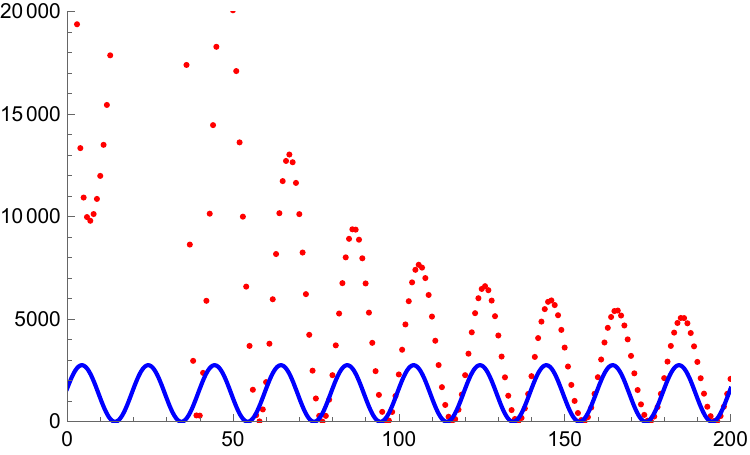}%
    \label{fig:cubic-E1-oscT1}}%
  \caption{The function $\imag \, \mc{E}(\nu,\hbar)$ evaluated on the
    sequence $\hbar^{(1)}_n$ for (a) $\nu=1/2$ and (b) $\nu=3/2$ with
    $\xi = 10$.  The red dots are numerical results, and the blue curve
    is the prediction from resurgence analysis.}
  \label{fig:cubic-oscT1}
\end{figure}

\begin{figure}
  \centering
  \subfloat[$\nu=1/2$]{\includegraphics[height=3cm]{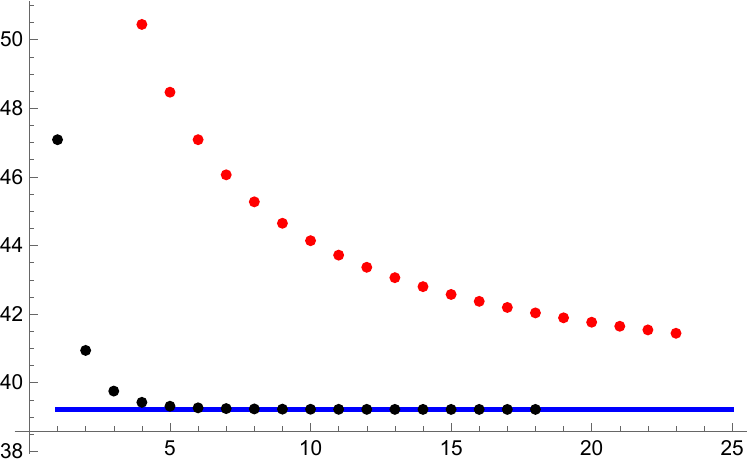}%
    \label{fig:cubic-E0-oscT2}}%
  \hspace{12ex}
  \subfloat[$\nu=3/2$]{\includegraphics[height=3cm]{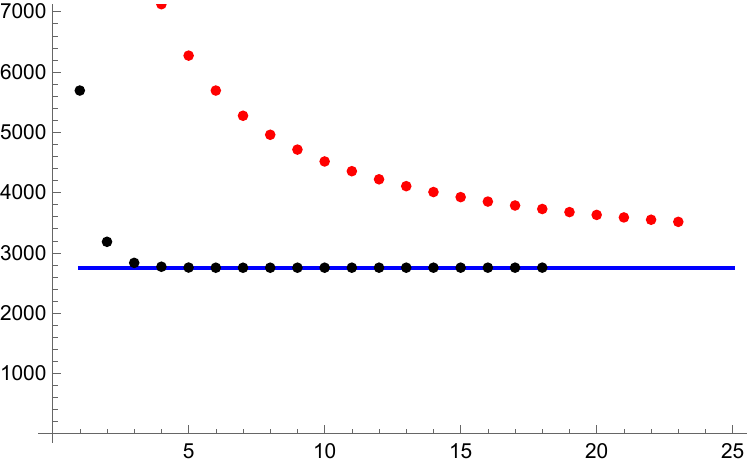}%
    \label{fig:cubic-E1-oscT2}}%
  \caption{The function $\imag \, \mc{E}(\nu,\hbar)$ evaluated on the
    sequence $\hbar^{(2)}_n$ for (a) $\nu=1/2$ and (b) $\nu=3/2$ with
    $\xi = 10$.  The red dots are numerical results, the black dots are
    results after applying the Richardson transform to speed up the
    convergence, and the blue line is the asymptotic value predicted
    from resurgence analysis.}
  \label{fig:cubic-oscT2}
\end{figure}

The formula \eqref{imr-ground} can also be tested directly by the
method discussed at the end of section~\ref{sec:comp}.  Let us
evaluate numerically the function
\begin{equation}
  \imag\,
  \mc{E}(\nu,\hbar):=-(2/\hbar)^{1-\nu}(\pi/2)^{1/2}\Gamma(\nu+1/2)
  \re^{\CS_\IR/\hbar-2\nu\CA_\IR}\imag \, E(\nu,\hbar)
\end{equation}
on a sequence of $\hbar$,
\begin{equation}
  \hbar^{(1)}_n = -10 \CS_{\CI}/(\pi n),\quad n=1,2,3,\ldots.
\end{equation}
Eq.~\eqref{imr-ground} predicts that the value of
$\imag\,\mc{E}(\nu,\hbar^{(1)}_n)$ should oscillate following the pattern
\begin{equation}\label{eq:cubic-E0-oscT1}
  \imag\,\mc{E}(\nu,\hbar^{(1)}_n) \approx \omega^{1+\nu}(x_\star - x_0)^{2\nu}
  (1+\cos(-(\pi/10) n-2\nu\CA_\CI)),
\end{equation}
in the large $n$ limit.  As illustrated by Fig.~\ref{fig:cubic-oscT1},
we find that this is indeed the case for both $\nu=1/2$ and $\nu=3/2$:
not only do the value of $\imag\,E(\nu,\hbar^{(1)}_n)$ and the
prediction from the resurgence analysis \eqref{eq:cubic-E0-oscT1}
oscillate locked in phase, the amplitude of oscillation of
$\imag\,\mc{E}(\nu,\hbar^{(1)}_n)$ also asymptotes to that of
\eqref{eq:cubic-E0-oscT1} in the large $n$ limit.  To test with higher
numerical precision the latter statement, as the prefactor in
\eqref{eq:cubic-E0-oscT1} is crucial, we evaluate
$\imag\,\mc{E}(\nu,\hbar)$ on a second sequence of $\hbar$ \jjj{Corrected}
\begin{equation}
  \hbar^{(2)}_n =-\CS_{\CI}/(2\pi n - 2\nu\CA_\CI),\quad n=1,2,3,\ldots,
\end{equation}
always at the crests of the oscillation, and its value should
asymptote to the maximum value
\begin{equation}
  \imag\,\mc{E}(\nu,\hbar^{(2)}_n) \approx \omega^{1+\nu}(x_\star - x_0)^{2\nu},
\end{equation}
which is indeed verified as shown in Fig.~\ref{fig:cubic-oscT2}.

The formula \eqref{imr-ground} shows that indeed the complexification
of the instantons lead to suppression of tunneling, as in the double
well. However, the precise pattern for the vanishing of the imaginary
part of the energy is due to the instanton with complex action, its
conjugate, and an additional instanton with real action. The latter
does not appear if we just use the naif complexification approach of
\cite{blgzj}. It is also not obvious how to understand this real
instanton in terms of complexified solutions to the EOM. In this case,
however, the resurgent approach gives us a precise understanding of
the non-perturbative corrections that have to be taken into account.

There is yet another perspective to understand the change of
singularity structure in the Borel plane, and therefore the phase
transition in the quantum mechanical model. The spectrum of BPS states
of $\CN=2$ super Yang--Mills theory is known to undergo wall-crossing
at special loci in moduli space \cite{sw,fb}, and the monopole points
typically belong to this loci. It was shown in \cite{ggm} that this
spectrum governs the singularity structure of the quantum periods
associated to the quantum SW curve. Therefore, the phase structure in
deformed quantum mechanics must be due ultimately to the wall-crossing
in the $\CN=2$, $SU(3)$ Yang--Mills spectrum. The splitting of
singularities we observe seems to be closely related to the appearance
of a tower of particles in the weakly coupled phase of the
${\cal N}=2$ theory, as it happens in the $SU(2)$ case.  It would be
interesting to make this more explicit.

\begin{table}
  \centering
  \resizebox{\linewidth}{!}{
  \begin{tabular}{*{3}{>{$}l<{$}}}\toprule
    \nu=1/2
    & \hbar = 3/2
    & \hbar = 1 \\
    \midrule
    \mr{S}^{(+)}E^{\text{pert}}(\hbar)
    & 2.158989556195731030 - \ri\,2.48522\times 10^{-13}
    & 1.377257148742246102603174 - \ri\,3.7969395719\times 10^{-14}\\
    E^{\text{num}}(\hbar)
    & 2.158989556195731029 - \ri\,2.48572\times 10^{-13}
    & 1.377257148742246102603189 - \ri\,3.7969395749\times 10^{-14}\\
    E^{(1)}(\hbar)
    & \CO(10^{-10})
    & \CO(10^{-15})\\
    \midrule
    \nu=3/2
    & \hbar = 3/2
    & \hbar = 1 \\
    \midrule
    \mr{S}^{(+)}E^{\text{pert}}(\hbar)
    & 6.8228767656714 - \ri\,3.23338\times 10^{-8}
    & 4.33475553996766479544 - \ri\,5.12436612\times 10^{-12}
    \\
    E^{\text{num}}(\hbar)
    & 6.8228767656751 - \ri\,3.23348\times 10^{-8}
    & 4.33475553996766479532 - \ri\,5.12436607\times 10^{-12}
    \\
    E^{(1)}(\hbar)
    & \CO(10^{-9})
    & \CO(10^{-14})\\
    \bottomrule
  \end{tabular}}
\caption{Comparison of the (upper) lateral Borel resummed perturbative
  energy series $\mr{S}^{(+)}E^{\text{pert}}(\nu,\hbar)$ and the
  numerical evaluation of the eigen-energy $E^{\text{num}}(\nu,\hbar)$
  at both $\nu=1/2$ and $\nu=3/2$ for $\xi=12$.  The convention is the
  same as in Tab.~\ref{tab:Eisum-hsmall}, except that for Borel
  resummation, the perturbative series is truncated to 400 terms for
  $\nu=3/2$, and for numerical evaluation the Hamiltonian matrix is
  truncatd to rank 250.}
  \label{tab:Eisum-hlarge}
\end{table}

\begin{figure}
  \centering
  \subfloat[$\xi=1/2$]{\includegraphics[height=4cm]{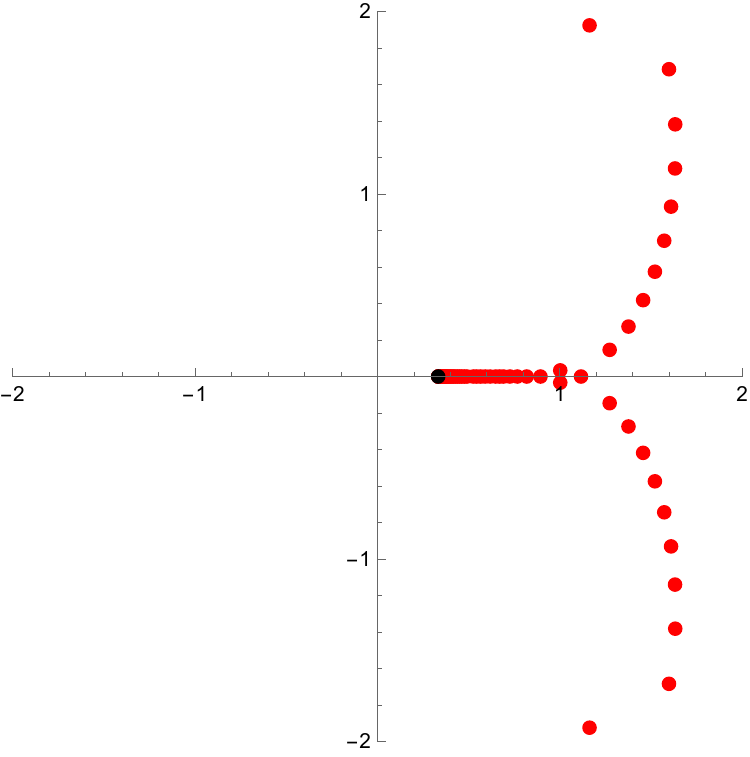}}\hspace{14ex}
  \subfloat[$\xi=8$]{\includegraphics[height=4cm]{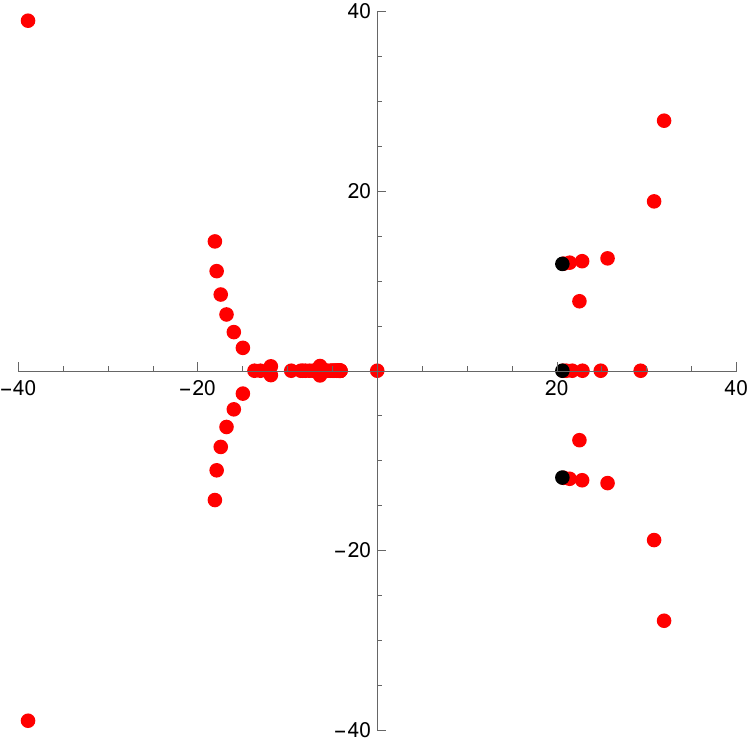}}
  \caption{The Borel plane of the perturbative energy series of the
    double-well model at $\nu=1/2$ for both $\xi<\xi_c$ (left) and $\xi>\xi_c$
    (right).  The red dots are position of poles of Pad\'e approximant
    of the Borel transform of the perturbative series truncated up to
    150 terms.  The black dots are the position of the branch points,
    which are located at $2\CS$ in the left plot, and at
    $2\CS_\IR, 2\CS_\IR\pm 2\ri \CS_\CI$ in the rigth plot.}
  \label{fig:dw-brl}
\end{figure}

We finally comment on the double-well model.  The resurgence analysis
can also be performed on the perturbative energy series for this
model, which is calculated to be
\begin{equation}
  E^{\text{pert}}(\nu,\hbar) = \sqrt{2\xi}\nu \hbar
  +
  \left(\left(-\frac{1}{8\xi}+\frac{\xi}{32}\right)
    +\left(-\frac{3}{2\xi}+\frac{\xi}{8}\right)\nu^2\right)
  \hbar^2
  + \CO(\hbar^3).
\end{equation}
However, as shown in Figs.~\ref{fig:dw-brl}, the leading singularities
of the Borel transform of the perturbative energy on the right half of
the Borel plane are located at $2\CS$ or $2\CS_\IR$ instead of at
$\CS$, indicating that it completely misses the one-instanton
corrections to the energy, just like in the ordinary quantum
mechanical model with a double-well potential (see e.g. \cite{du} for
a discussion.)

\subsection{Comparison to the exact quantization conditions}
One of the original reasons to consider the deformation of quantum mechanics 
studied in this paper was to have a testing 
ground and a simpler example of the TS/ST duality of \cite{ghm,cgm}, which 
provides an explicit, exact solution of certain spectral problems in the quantum 
mechanics of Weyl operators via topological string theory. The Hamiltonian of 
deformed quantum mechanics turns out to be a special limit of the Weyl-like 
Hamiltonians considered in \cite{ghm,cgm}. 

The EQCs in \cite{gm-dqm} are expressed in terms of a special resummation 
of the quantum periods on the spectral curve 
\be
\label{SWN-curve}
2\Lambda^N \cosh(p)+W_N(x)=0,
\ee
 where $W_N(x)$ was defined in (\ref{wnx}), and the Liouville form is $p(x) \rd x$. As mentioned in section \ref{sec:dQM}, 
 these periods can be obtained by using the so-called NS limit \cite{ns} of the instanton partition function \cite{lns,nn} 
 of the underlying $\CN=2$ super Yang--Mills theory. The details are not important for our purposes, since we will only need the classical limit of these periods. A more detailed description of the full quantum story can be found in \cite{gm-dqm}. There are two types of quantum periods in this game, 
 the so-called $A$ and $B$ periods, which correspond to a choice of $A$ and $B$ cycles in the curve (\ref{SWN-curve}). 
 We will denote them respectively by 
 \be
a_i (\{ \xi\}; \hbar), \quad \phi_i (\{ \xi\} ; \hbar), \qquad i=1, \cdots, N. 
 \ee
 The classical limit of the $\phi_i (\{ \xi\} ; \hbar)$ as $\hbar \to 0$, which we will denote as $\phi^{(0)}_i (\{ \xi\} ; \hbar)$, are the classical 
 periods which appear in the EBK quantization conditions of the Toda lattice (see e.g. \cite{matsuyama} or section 2.12 of \cite{aqm}). 
The EQCs of \cite{gm-dqm} are given by a single functional constraint 
on the quantum periods. One important remark is that the quantum periods 
appearing in the EQC are defined by Yang--Mills instanton partition sums, 
which do not converge everywhere in moduli space, and this might lead to limits in the 
applicability of these conditions, as already noted in \cite{gm-dqm}. Most relevant to 
our problem, when $\hbar=0$, the instanton sums 
only converge in the weak coupling region, and for $\hbar$ small we do not 
expect them to converge deep inside the strong coupling region. Therefore, in order to obtain the 
asymptotics of the non-perturbative 
effects in the strong coupling phase, $\xi<\xi_c$, $E =\CO(\hbar)$, and $\hbar \to 0$, 
one would have to make an extrapolation 
of the quantum periods beyond its region of convergence, a problem we will not address here.

Let us now write down the EQCs found in \cite{gm-dqm}. In the case of the double-well, 
symmetry reasons impose the following relationship between the periods: $a_1=a_3$, $\phi_3=\phi_2$. 
The quantization conditions read
\be\label{symqc2}
\cos\left( {\phi_2\over 2 \hbar}\right)=  \epsilon \CV \sin \left( {\phi_1 + \phi_2 \over 2\hbar}  \right), \qquad \epsilon=\pm (-1)^k, 
\ee
where the sign $\pm (-1)^k$ corresponds to the energy level $E_{2k}$, $E_{2k+1}$, respectively, and $\CV$ is given by 
\be
\label{np-factor}
\CV= \re^{- \pi a_2/\hbar} \frac{1-\re^{-2 \pi  a_1/\hbar }}{1-\re^{-2 \pi  (a_1+a_2)/\hbar }}.
\ee
We note that, in order to have an easier match with the conventions in this paper, we have redefined $\phi_i \rightarrow 
\phi_i/\hbar$, and changed the sign of $\phi_2$ in the equations of \cite{gm-dqm}. 

\begin{figure}
  \centering
\includegraphics[height=4.4cm]{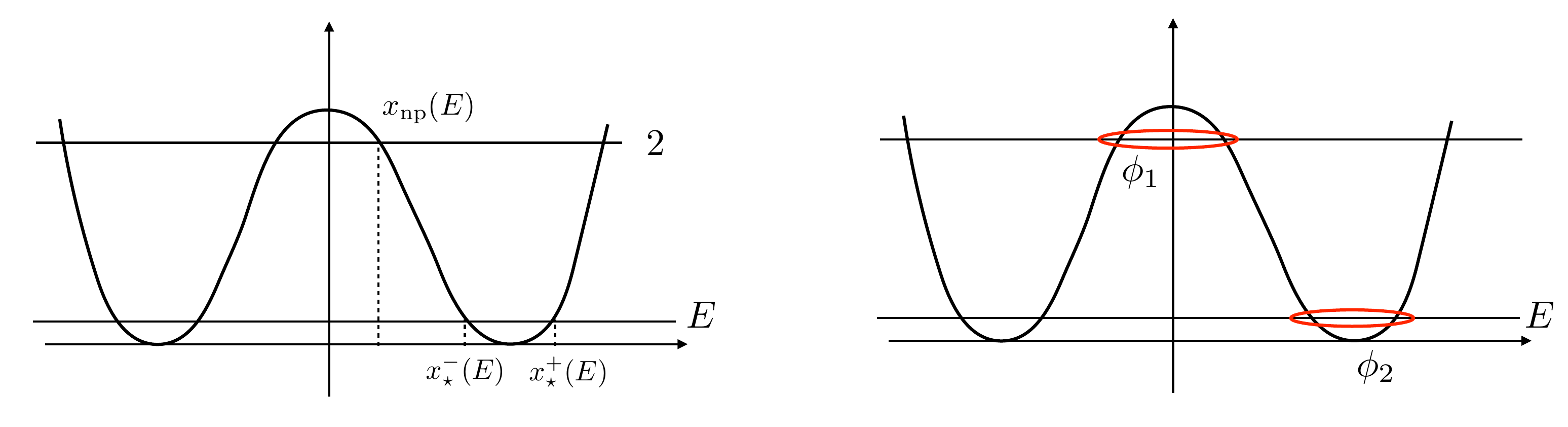}
  \caption{The double-well potential in the weak coupling phase. In the figure on the left we indicate the points $x_\star^\pm (E)$ and $x_{\rm np}(E)$. On the figure on the right, we show the cycles that lead to the periods $\phi_{1,2}$ in the EQC.}
  \label{fig:dw-cycles}
\end{figure}

To understand the EQC in detail, we have to be more precise about the quantum periods appearing in it. Let us first define the points $x_\star^\pm(E)$ as the positive turning points of $u (x)$ for an energy $E$:
\be
u\left( x_\star^\pm(E) \right)=E. 
\ee
When $E=0$, they become the critical point $x_\star$. They are shown in \figref{fig:dw-cycles}. The classical limits $\phi_i^{(0)}(E)$ of the quantum periods $\phi_i$ are given by 
\be
\label{class-limit}
\ba
\phi_1^{(0)}&= 2 \int_{x^-_{\rm np}(E)}^{x^+_{\rm np}(E)} \cosh^{-1} \left( u_4(z)-1- E \right) \rd z,\\
\phi_2^{(0)}&= 2 \int_{x^-_\star(E)}^{x^+_\star (E)} \cosh^{-1} \left( 1- u_4(z) + E \right) \rd z.  
\ea
\ee
Therefore, $\phi_2^{(0)}$ is the perturbative WKB period around the critical point, while $\phi_1^{(0)}$ is the period which appears in the weak coupling phase (the periods also depend on $\xi$, but we have not indicated this dependence explicitly). The quantity $\CV$ is non-perturbative, and involves tunneling periods. By using the instanton expressions in \cite{gm-dqm} it can be seen that, in this phase, $a_{1,2}>0$, and in addition 
\be
\pi a_2= \Pi_4 (E)+ \CO(\hbar^2), 
\ee
where
\be
\label{PiE}
\Pi_4(E)=2 \int_{x_{\rm np}(E) }^{x^{-}_\star (E) } \cos^{-1} \left( 1- u_4(z) + E \right) \rd z + 2 \pi  x_{\rm np}(E) 
\ee
is the tunneling cycle from $-x^-_\star(E)$ to $x^-_\star(E)$. 
Therefore, at leading order in both $\hbar$ and exponentially small, non-perturbative terms, we have
\be
\CV\approx  \re^{-\Pi_4(E)/\hbar}. 
\ee

We can now analyze the quantization condition (\ref{symqc2}) in some detail. The first observation is that, 
if we neglect non-perturbative corrections, we have 
\be
\label{bs-cond}
\cos\left( {\phi_2\over 2 \hbar}\right)=0 \Rightarrow \phi_2= 2 \pi \hbar \nu, 
\ee
where $\nu$ was introduced in (\ref{nu-def}). This is the standard all-orders 
Bohr--Sommerfeld quantization condition for this problem. For $E, \hbar \to 0$, 
one can see from an explicit computation of the classical period that (\ref{bs-cond}) becomes (\ref{Ehbar}). A second observation is that the EQC has a special solution of the form 
\be
\label{toda-4}
\phi_1= 2 \pi \hbar \left( s+ {1\over 2} \right), \quad \phi_2= 2 \pi \hbar \left( n+ {1\over 2} \right)
\ee
where $s$ is a non-negative integer, and $\phi_2$ is as in (\ref{bs-cond}). Since we 
have two conditions, this fixes the values of both $E$ and $\xi$. The solution is a discrete 
set of values for these two quantities, which are precisely the Toda lattice points mentioned 
in section \ref{sec:dQM}. Indeed, the conditions (\ref{toda-4}) are a particular case of the EQCs for the periodic Toda 
lattice obtained in \cite{ns} for $u(x)$ even, so these values agree with the spectrum of the 
corresponding conserved charges for this integrable system, in the case $N=4$ (the general solution for the 
$N=4$ periodic Toda lattice appears in the quantum-mechanical model with an arbitrary quartic potential.) 

Let us now compute the energy gap as predicted from (\ref{symqc2}). To do this, we 
use the idea of \cite{alvarez}: we perturb the Bohr--Sommerfeld quantization condition (\ref{bs-cond}) as 
\be
\label{pbs}
\nu= n+{1\over 2} + \Delta \nu, 
\ee
where $\Delta \nu$ is non-perturbative. One immediately finds, 
\be
\Delta \nu  \approx  - {\epsilon  \over \pi} \CV \cos\left( {\phi_1 \over 2 \hbar} \right)+ \CO(\CV^2), 
\ee
and from here we obtain the energy gap at leading non-perturbative order: 
\be
\Delta E (\nu)=-{2 \over \pi}  {\rd E \over \rd \nu} \re^{-\pi a_2 /\hbar} \cos\left( {\phi_1 \over 2 \hbar} \right). 
\label{split-EQC}
\ee
To make contact with the formula (\ref{eq:complex-dw}) we have to take the limit $E, \hbar \rightarrow 0$ 
of this expression. In the case of $\phi_{1,2}$, we find immediately, after using (\ref{Ehbar}), 
\be
{\phi_1\over 2 \hbar} ={\CS_{\CI} \over \hbar}- 2 \nu \CA_{\CI}+\CO(\hbar^2). 
\ee
 The limit $E \rightarrow 0$ of (\ref{PiE}) has a logarithmic singularity which has to be extracted as we have done before in e.g. (\ref{PiE0}). In fact, we find the same expression but involving just the real part of the action and $\CA$:
 \be
\label{PiE0b}
\Pi_4(E) =\CS_{\IR}+ {E\over \omega} \left( \log (2 E)-E \right)-{2E \over \omega} \left( \CA_{\CR} + \log(2 \omega (x_\star -x_0)) \right) + \cdots,
\ee
By using all the results, it is easy to see that the small $\hbar$ limit of (\ref{split-EQC}) reproduces exactly (\ref{eq:complex-dw}). 

As noted already in \cite{gm-dqm}, the energy gap vanishes and
tunneling is suppressed exactly at the points (\ref{toda-4}). The
formula (\ref{eq:complex-dw}) gives a small $\hbar$ approximation to
the gap, where in addition only the leading non-perturbative effect
has been taken into account. The points where tunneling is suppressed
can be found, at small $\hbar$, by determining the values of $\xi$ at
which the $\cos$ term vanishes, for each energy level $n$. It is an
easy exercise to calculate this from the explicit expressions of
$\CS_\CI$, $\CA_\CI$. In particular, one has to evaluate $\CA_\CI$ at the
critical point $\xi=\xi_c$. In this limit, the integral defining $\CA_\CI$
has contributions from the harmonic oscillator at $x=x_\star$, with
frequency $\omega_c= \omega (\xi_c)$, and from the inverted harmonic
oscillator at the local maximum at $x=0$, with frequency
$\omega_c' =2$. One finds
\be
\label{gen-aic}
\CA_\CI(\xi_c)= - \pi {\omega_c' \over \omega}=- {\sqrt{2} \over 2} \pi, 
\ee
and this leads to the expression
\be
\xi_{\rm ST}= 4+ 2\left( s+ {\sqrt{2} \over 2} \left(n +{1\over 2} \right)+{1\over 2} \right) \hbar+ \CO(\hbar^2), 
\ee
where $s$ corresponds to the quantum number appearing in the quantization condition of the 
Toda lattice (\ref{toda-4}). The subscript ${\rm ST}$ refers to suppression of tunneling. 

\begin{figure}
  \centering
\includegraphics[height=4.6cm]{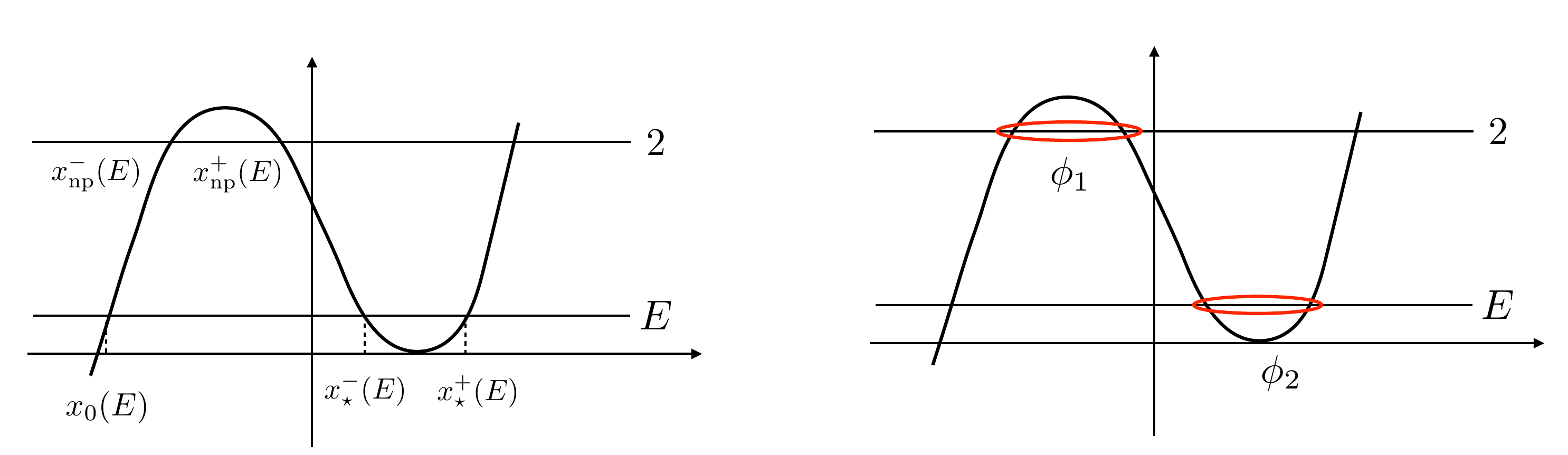}
  \caption{The cubic potential in the weak coupling phase. In the figure on the left we indicate the points $x^\pm_\star (E)$ and $x^\pm_{\rm np}(E)$. On the figure on the right, we show the cycles that lead to the periods $\phi_{1,2}$ appearing in the EQC.}
  \label{fig:cubic-cycles}
\end{figure}

Let us now consider the cubic potential. In this phase, the potential looks as depicted in the figure on the left in \figref{fig:cubic-cycles}. 
The EQC conditions read in this case 
\be
\label{n3eqc}
1+ {1 -\re^{-2 \pi a_1/\hbar} \over 1- \re^{-2 \pi a_2 /\hbar}}\re^{\ri \phi_2/\hbar} - 
{\re^{-2 \pi a_2/\hbar}  -\re^{-2 \pi a_1/\hbar} \over 1- \re^{-2 \pi a_2 /\hbar}}  \re^{\ri \phi_1/\hbar}=0. 
\ee
We have rearranged the periods $a_i$ w.r.t. what is written down in \cite{gm-dqm}: the period $a_1$ there is $-a_1$ here, and the period $-a_1-a_2$ there is $a_2$ here. With the new notation, we have in this phase $a_{1}>a_2>0$. The $B$-periods $\phi_1$, $\phi_2$ correspond to the cycles depicted in the figure on the right in \figref{fig:cubic-cycles}. Their classical limits are given by the same equations as in (\ref{class-limit}), 
but with $u_3(z)$ instead of $u_4(z)$. 
In addition, we have 
\be
2 \pi a_2 = \Pi_3(E)+ \CO(\hbar)
\ee
where
\be
\Pi_3(E)=2\left(  \int_{x_0(E)}^{x^-_{\rm np}(E)}+  \int_{x^+_{\rm np}(E)}^{x^-_{\star}(E)} \right) \cos^{-1}\left(1- u_3(z)  + E  \right)\rd z + 2\pi \left(x^+_{\rm np}(E)-x^-_{\rm np}(E)\right). 
\ee
Let us now analyze the EQC (\ref{n3eqc}). When all non-perturbative corrections 
involving $a$ periods are neglected, we find again the all-orders Bohr--Sommerfeld quantization condition 
(\ref{bs-cond}), and it is easy to see that the analogue of (\ref{toda-4}) solves the EQC exactly. These two 
equations give again the spectrum of the quantum periodic Toda lattice with $N=3$. Let us now compute 
the imaginary part of the energy from (\ref{n3eqc}) by considering only the leading non-perturbative correction. 
Since $a_1>a_2$, this correction is due to $\re^{-2 \pi a_2/\hbar}$. Like before, and as in 
\cite{alvarez-casares2}, we consider a perturbed Bohr--Sommerfeld solution (\ref{pbs}) and we find 
\be
{\rm Im}\, \Delta \nu \approx {1\over 2 \pi} \left( 1+ \cos\left({\phi_1 \over \hbar} \right) \right) \re^{-2 \pi a_2/\hbar}, 
\ee
which leads in turn to
\be
\label{imwp}
{\rm Im}\, E( \nu) \approx -{1\over 2 \pi} {\partial E \over \partial \nu} \left( 1+ \cos\left({\phi_1 \over \hbar} \right) \right) \re^{-2 \pi a_2/\hbar}. 
\ee
Here we have taken into account that the EQCs of \cite{gm-dqm} give the imaginary 
parts of the energies of the resonances with a $+$ sign, which is the opposite to our convention. 
Since in the cubic case we have 
\be
{\phi_1\over  \hbar} ={\CS_{\CI} \over \hbar}- 2 \nu \CA_{\CI}+\CO(\hbar^2), 
\ee
one finds that (\ref{imwp}), in the $\hbar \to0$ limit, reproduces (\ref{imr-ground}), 
%
%
in precise agreement with the resurgent analysis. Like in the double-well potential, we can use this 
formula to find the small $\hbar$ expression for the values of $\xi$ at which tunneling is suppressed. 
The evaluation of $\CA_I$ at $\xi=\xi_c$ is given by the same formula (\ref{gen-aic}), but now 
$\omega'_c= \omega_c={\sqrt{3}}$. We find in the end,  
\be
\xi_{\rm ST}= 3+ {\sqrt{3}} (s+n+1)\hbar+ \CO(\hbar^2). 
\ee

\section{The critical theories}
\label{sec:crit}

\begin{figure}
  \centering
\includegraphics[height=3.7cm]{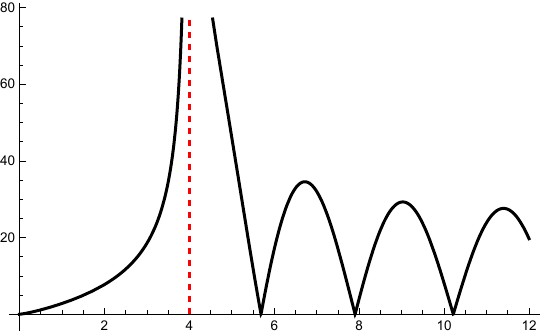}\qquad \qquad \qquad   \includegraphics[height=3.7cm]{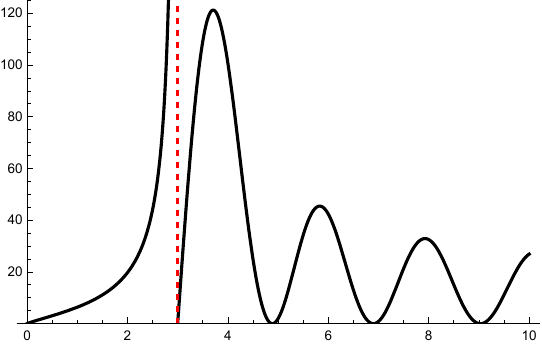}
  \caption{The behavior of the one-loop prefactor for the energy gap of the double well (left) and for the imaginary part of the cubic well (right), 
  as a function of $\xi$, and we have set $\hbar=1$ for convenience. The red vertical lines are at $\xi=\xi_c$, where the functions are singular. When $\xi>\xi_c$ they have an oscillatory behavior and tunneling is suppressed at special points. }
  \label{fig:prefactors}
\end{figure}
The small $\hbar$ asymptotics 
of the energy gap and the imaginary part of the energies, as a function of $\xi$, 
is discontinuous as we approach the critical value of the parameter $\xi=\xi_c$, independently 
of the value of $\hbar$. In the case of the double-well potential, the gap diverges as we approach $\xi=\xi_c$ from 
both the strong and the weak coupling region, since the function $\CA$ in the one-loop prefactor 
diverges at the critical point. In the cubic well, the imaginary part of the energy diverges 
as we approach $\xi=\xi_c$ from the 
strong coupling region, and it vanishes in the weak coupling region due to the oscillatory factor, see \figref{fig:prefactors} for a graphical depiction of this behavior. Of course, in the 
quantum-mechanical spectral problem, the gap and the imaginary part of the energy are finite when $\xi=\xi_c$ and for 
any $\hbar>0$, so the discontinuity indicates rather a change of asymptotic behavior 
in the critical regime. 

To find the critical behavior of the instanton amplitudes, we look at the WKB periods 
in the strong coupling phase, when $\xi<\xi_c$, and 
see what happens as we approach the critical point. Our derivation of the 
$\hbar \to 0$ asymptotics in section \ref{subsec-inst-wkb} relied 
crucially on the expansion (\ref{PiE0}) of the tunneling period near $E=0$. 
However, when $\xi=\xi_c$, this expansion is no longer valid. The reason is that there 
is a contribution to the logarithm coming from the inverted harmonic oscillator around the local maximum where $u(x)=2$. Let us first 
consider the cubic potential. At the critical point $\xi_c=3$, the correct small $E$ expansion of the tunneling WKB period is
\be
\Pi(E)= {3\over \omega_c} \left( E \log(2 E) -E \right)-2  {\cal D}_3+ \CO(E), 
\ee
where $\omega_c=  {\sqrt{3}}$ is the value of frequency at the critical point. The factor of $3/\omega_c$ can be understood as the contribution from one inverted oscillator at $x=-1$, with frequency $\omega_c'=\omega_c$, and half the contribution of the oscillator around $x=x_\star$ with the same frequency. 
The constant ${\cal D}_3$ can be computed by subtracting the divergent integrals carefully as in (\ref{split}). One finds  
\be
{\cal D}_3=\frac{3}{2} \sqrt{3} \log (3). 
\ee
This period then contributes to the non-perturbative amplitude as 
\be
\re^{- \Pi(E)/\hbar} =\left({1\over 2\hbar \omega_c} \right)^{3 \nu} \exp\left\{ -3 (\nu \log \nu-\nu) + 2 {\cal D}_3 \omega_c \nu+ \cdots  \right\}. 
\ee
As we mentioned in section \ref{subsec-inst-wkb}, a crucial ingredient to obtain the correct asymptotics for a fixed energy level involves incorporating the singular part of the all-orders WKB expansion and promoting the logarithmic terms in $\nu$ to (\ref{G-pref}). However, in this case, the appearance of an additional harmonic oscillator at the maximum changes this. The correct incorporation of the all-order corrections involves now  
\be
\label{prom}
\re^{-3 (\nu \log \nu-\nu)} \rightarrow \left[{\sqrt{2 \pi} \over \Gamma\left( \nu +{1\over 2} \right)} \right]^3. 
\ee
We have verified that this is the right correction at NLO in the WKB expansion, by using (\ref{p1}). We then obtain, 
\be
\label{ccw-nu}
{\rm Im}\, E (\nu) \approx -{1\over 4} { {\sqrt{2 \pi}} \over \Gamma^3 \left( \nu +{1\over 2} \right) } \left({1\over 2\hbar \omega_c} \right)^{3 \nu-1}  \re^{-\CS_c/\hbar+ 2{\cal D}_3 \omega_c \nu}, 
\ee
where $\CS_c$ is the value of the action at the critical point. In particular, when $\nu=1/2$ we obtain 
\be
{\rm Im}\, E_0 \approx -{1\over 4} \left({ \pi\over \hbar \omega_c} \right)^{1/2}  \re^{-\CS_c/\hbar+ {\cal D}_3 \omega_c }. 
\ee
Note that the one-loop correction scales now with $\hbar$ as $\hbar^{-1/2}$,
instead of $\hbar^{1/2}$, and the one-loop prefactor diverges as
$\hbar \to 0$.

\begin{figure}
  \centering
  \subfloat[$\nu=1/2$]{\includegraphics[height=3cm]{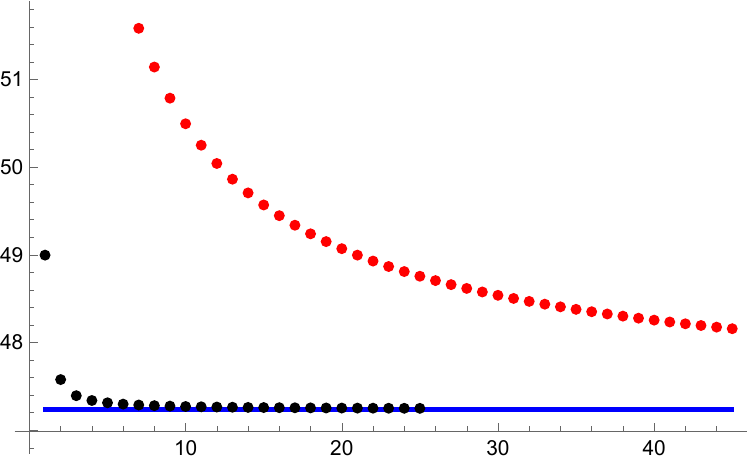}}\hspace{14ex}
  \subfloat[$\nu=3/2$]{\includegraphics[height=3cm]{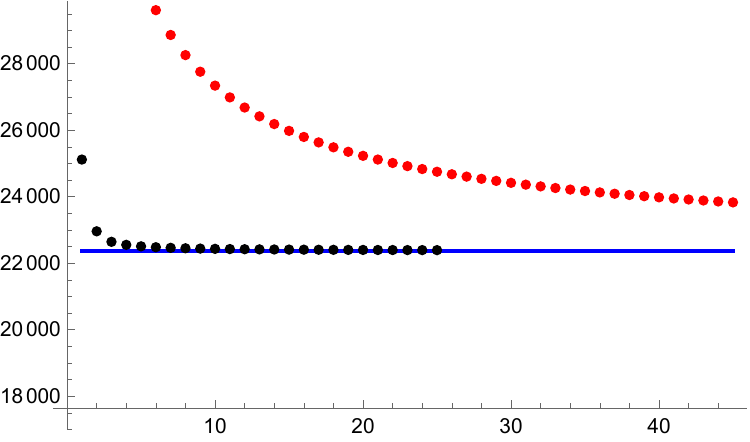}}
  \caption{The function $\imag \, \mr{E}(\nu,\hbar)$ evaluated on
    $\hbar = 1/n$, $n=1,2,3,\ldots$, for $\nu=1/2$ (left) and
    $\nu=3/2$ (right).  The red dots are numerical results, the black
    dots are results after applying the Richardson transform to speed
    up the convergence, and the blue line is the asymptotic value
    predicted by WKB calculation.}
  \label{fig:cubic-CRT}
\end{figure}

We can test \eqref{ccw-nu} numerically by calculating the function
\begin{equation}
  \imag\,\mr{E}(\nu,\hbar) := -\imag\,E(\nu,\hbar)\,\hbar^{3\nu-1} \re^{\CS_c/\hbar}
\end{equation}
on the sequence of $\hbar = 1/n$, $n=1,2,3,\ldots$, and it should have
the asymptotic behavior
\begin{equation}
  \imag\,\mr{E}(\nu,1/n) \approx {1\over 4} { {\sqrt{2 \pi}} \over
    \Gamma^3 \left( \nu +{1\over 2} \right) } \left({1\over 2
      \omega_c} \right)^{3 \nu-1}  \re^{2{\cal D}_3 \omega_c \nu} + \mc{O}(1/n),
\end{equation}
which is indeed the case, as shown in Fig.~\ref{fig:cubic-CRT}.

The analysis of the double-well potential is similar. The inverted oscillator has now a frequency $\omega_c'=2$, 
while $\omega_c= 2 {\sqrt{2}}$. The tunneling period has the expansion
\be
 \Pi(E) = \left( {1\over \omega_c} + {1 \over \omega'_c} \right)\left(E \log(2 E) -E \right)-2 {\cal D}_4+\cdots
\ee
where 
\be
{\cal D}_4 =\frac{1}{8} \left(8 \left(2+\sqrt{2}\right) \log (2)+\log \left(17-12 \sqrt{2}\right)-2
   \sqrt{2} \log \left(3+2 \sqrt{2}\right)\right). 
\ee
The contribution of the tunneling period is then
\be
\re^{- \Pi(E)/\hbar} \approx \left({1\over 2\hbar \omega_c} \right)^{ \nu \left(1+ {\omega_c \over \omega_c'} \right)} \re^{- \left(1+ {\omega_c \over \omega_c'} \right) (\nu \log \nu-\nu) + 2 {\cal D}_4 \omega_c \nu}. 
\ee
We have to find now the correct form of the all-order corrections. One indication is that the two harmonic oscillators at $x=0$ and $x={\sqrt{2}}$ contribute equally but with different frequencies, and the right prescription is 
\be
\exp\left\{  - \left(1+ {\omega_c \over \omega_c'} \right) (\nu \log \nu-\nu)  \right\} \rightarrow  {\sqrt{2 \pi} \over \Gamma\left( \nu +{1\over 2} \right)} {\sqrt{2 \pi} \over \Gamma\left( {\omega_c \over \omega_c'} \nu +{1\over 2} \right) } \exp\left[ {\omega_c 
\over \omega_c'}\nu  \log\left( {\omega_c \over \omega_c'} \right) \right], 
\ee
which can also be verified by looking at the NLO correction. 
We then find for the energy splitting
\be\label{delE-nu-CRT}
\Delta E (\nu)\approx \left({1\over 2\hbar \omega_c} \right)^{ \nu \left(1+ {\omega_c \over \omega_c'} \right)-1} {\re^{-\CS_c /\hbar} \over \Gamma\left( \nu +{1\over 2} \right)\Gamma\left( {\omega_c \over \omega_c'} \nu +{1\over 2} \right) }\exp\left[ {\omega_c 
\over \omega_c'}\nu  \log\left( {\omega_c \over \omega_c'} \right) + 2 {\cal D}_4 \omega_c \nu  \right] . 
\ee
For the ground state we have, 
\be
\Delta E_0 \approx \left({1\over 2 \hbar \omega_c} \right)^{ {\sqrt{2} -1 \over 2}} {\re^{-\CS_c/\hbar} \over \Gamma\left( {\sqrt{2} \over 2} +{1\over 2} \right) }\exp\left[{\sqrt{2} \over 4}    \log(2) +  {\cal D}_4 \omega_c  \right]. 
\ee
The prefactor in the energy gap scales now with $\hbar$ as
\be
\hbar ^{- {\sqrt{2} -1 \over 2}}. 
\ee
In general, both non-perturbative effects display an ``anomalous"
scaling as a function of $\hbar$, for all the levels.

\begin{figure}
  \centering
  \subfloat[$\nu=1/2$]{\includegraphics[height=3cm]{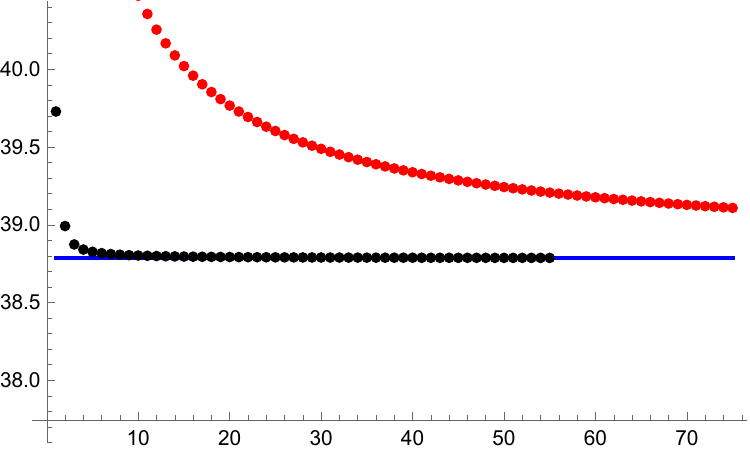}}\hspace{14ex}
  \subfloat[$\nu=3/2$]{\includegraphics[height=3cm]{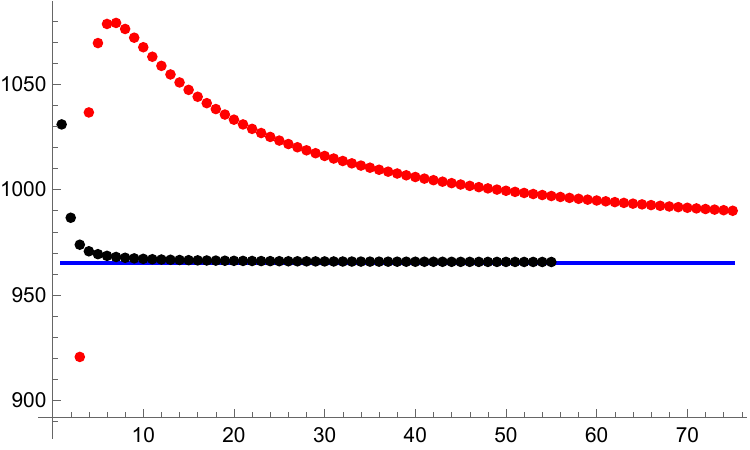}}
  \caption{The function $\Delta \mr{E}(\nu,\hbar)$ evaluated on
    $\hbar = 1/n$, $n=1,2,3,\ldots$, for $\nu=1/2$ (left) and
    $\nu=3/2$ (right).  The red dots are numerical results, the black
    dots are results after applying the Richardson transform to speed
    up the convergence, and the blue line is the asymptotic value
    predicted by WKB calculation.}
  \label{fig:dw-CRT}
\end{figure}

Similarly, we can test \eqref{delE-nu-CRT} numerically by calculating
the function
\begin{equation}
  \Delta \mr{E}(\nu,\hbar) := \Delta E(\nu,\hbar)\,\hbar^{\left(1+\frac{\omega_c}{\omega'_c}\right)\nu-1} \re^{\CS_c/\hbar}
\end{equation}
on the sequence of $\hbar = 1/n$, $n=1,2,3,\ldots$, and it should have
the asymptotic behavior
\begin{equation}
  \Delta \mr{E}(\nu,1/n) \approx  { \left({ 2 \omega_c} \right)^{ 1 - \nu \left(1+ {\omega_c \over \omega_c'} \right)}
    \over \Gamma\left( \nu +{1\over 2} \right)\Gamma\left( {\omega_c \over \omega_c'} \nu +{1\over 2} \right) }\exp\left[ {\omega_c 
      \over \omega_c'}\nu  \log\left( {\omega_c \over \omega_c'} \right) + 2 {\cal D}_4 \omega_c \nu  \right]  + \mc{O}(1/n),
\end{equation}
which is indeed the case, as shown in Figs.~\ref{fig:dw-CRT}.

In this derivation we have assumed that the behavior at the critical
point is obtained by extrapolation from the strong coupling phase, and
not from the weak coupling phase, which would give a different
result. However, the detailed numerical tests show that the formulae
above describe very precisely the small $\hbar$ asymptotics of both
the imaginary part of the energy in the cubic well and the energy gap
in the double well. This confirms our assumption, but we lack a
precise justification for this. Note that, from the point of view of
the underlying ${\cal N}=2$ super Yang--Mills theory, the critical
theories occur precisely at the monopole point, at the boundary
between the two phases, and at point where wall-crossing takes
place. The fact that the strong coupling spectrum is the one relevant
to understand the quantum periods exactly at this point is reminiscent
of a related observation in \cite{marino-schwick}.

\section{Conclusions}
\label{sec:con}
In this paper we have analyzed the behavior of the deformed
quantum-mechanical models introduced in \cite{gm-dqm} from the point
of view of instanton physics. Perhaps the most interesting aspect of
these models is the suppression of quantum tunneling at special points
in the moduli space, the so-called Toda lattice points. The instanton
analysis that we have performed uncovers the physics behind tunneling
suppression and unveils a rich phase structure: there is a strong
coupling phase with a conventional quantum-mechanical behavior, and a
weal coupling phase where the instantons become complex and lead to
suppression of tunneling. The two phases are separated by a critical
point where the non-perturbative effects have an ``anomalous"
scaling. In addition, the phase structure can be regarded as a
physical consequence of wall-crossing in the BPS spectrum of the
underlying ${\cal N}=2$ theories.

From the purely quantum-mechanical point of view, the mechanism for
suppression of tunneling in our example involves complex instantons
and occurs at special points in parameter space, as in \cite{garg,
  fefferman}. In contrast to these scenarios, it applies to a simpler,
one-dimensional system, and the mechanism for complexification of the
instantons is different and more intrinsic, since in \cite{garg,
  fefferman} suppression of tunneling requires an external magnetic
field. We should also point out that, in conventional quantum
mechanics, changes in the parameters of the potential lead to
wall-crossing and to different resurgent structures and quantization
conditions, as unveiled in the pioneering works of
\cite{voros-quartic, ddpham,DDP93}. However, the change in resurgent
structures in that case is triggered typically by a drastic change in
the classical potential, and by a rearrangement of classically allowed
and forbidden zones. In the deformed quantum mechanical models that we
studied in this paper, the nature of the classically forbidden region
for motion in real time remains the same for all values of the control
parameter $\xi$, but the motion in Euclidean time does change and
alters the instanton structure.

Although the wall-crossing phenomenon is clear from a resurgent
analysis of the perturbative series, it would be interesting to
understand it more directly, from the point of view of the $SU(N)$
Yang--Mills theories, and in particular in the $SU(3)$ case, where it
is richer. The connection between the wall-crossing in the spectrum
and the wall-crossing in the resurgent structure studied here is
however not completely direct. First, one has to connect the BPS
spectrum to the resurgent structure of the WKB periods, similarly to
what was done in \cite{ggm}, and then consider the perturbative
series, which themselves involve a limiting behavior of the WKB
periods akin to what was studied in \cite{marino-schwick}.  The
analysis of the $SU(3)$ spectrum in e.g.  \cite{longhi,yan} could be a
good starting point.

Another relationship that deserves further scrutiny is the connection
to the EQCs found in \cite{gm-dqm}.  We have found that these
conditions describe correctly the small $\hbar$ asymptotics of the
non-perturbative effects for the low-lying eigenstates, but only in
the weak coupling phase. The strong coupling phase is governed by a
different structure. The reason is that the EQCs are expressed in
terms of instanton partition sums, which are not expected to converge
in the strong-coupling region $\xi<\xi_c$, $E=\CO(\hbar)$ and
$\hbar \to 0$. An important problem is whether the existence of such a
phase structure extends to other examples. The TS/ST correspondence
\cite{ghm,cgm, wzh} provides EQCs for all the spectral problems
associated to toric Calabi--Yau (CY) manifolds, and the deformed
quantum-mechanical model of this paper is just one example among
others. The EQCs obtained in the TS/ST correspondence are expressed in
terms of topological string partition functions at large radius, which
also have a limited region of convergence. In many cases, in
particular in CYs with a single modulus, this region arguably covers
all possible values of the spectrum for real positive $\hbar$, and no
phase structure seems to occur.  It would be interesting to re-examine
these EQCs, specially in the multi-modulus case studied in \cite{cgm},
to see whether similar phase structures can occur, due to a breakdown
in the convergence of the partition functions involved in the EQCs. At
the same time, we should emphasize that the EQCs of \cite{gm-dqm} give
very explicit and detailed predictions on the non-perturbative effects
appearing in the weakly-coupled phase. In this paper we have only
probed the dominant ones, but the conjectural exact formulae
(\ref{symqc2}) and (\ref{n3eqc}) contain in principle all of them, and
therefore information about the full resurgent structure (in comparing
both, one has to take into account though that the quantum periods
appearing in the instanton sums and the ones in the exact WKB method
might be different, as shown in \cite{ggm}).

Although we have focused on the most emblematic cases, the double well
and the cubic potential, it would be interesting to find a general
pattern for potentials of higher degree, where these phenomena also
occur.  Recently, based on the open string version of the TS/ST
correspondence \cite{mz-wave,mz-wave2,gm1, gm2}, exact expressions for
the eigenfunctions of the deformed quantum-mechanical models of
\cite{gm-dqm} have been found in \cite{gmp}. It would be very
interesting to examine the phase structure found in this paper from
the point of view of these wavefunctions.

Another interesting problem is to develop more rigorous methods to
study the type of difference equations that appear in deformed quantum
mechanics, and perhaps prove rigorously some of the results in
\cite{gm-dqm} and in this paper. For example, the asymptotic behavior
of the energy gap in double-well systems has been derived rigorously
(see \cite{simon} for a classic paper and \cite{fefferman} for more
recent results). The phase structure found in this paper, and in
particular the ``anomalous" scaling at the critical point, seem ideal
laboratories to extend the spectral theory of the energy gap.  In
another direction, developing the exact WKB approach to these
equations along the lines of \cite{delmonte, baldino} might lead to a
derivation of both the EQCs of \cite{gm-dqm} and to a derivation of
the phase structure uncovered here. Recent results on wavefunctions of
the Baxter operator of the Toda lattice \cite{teschner} might also be
useful in that respect.

 \section*{Acknowledgements}
 M.M. would like to thank Slava Rychkov for reigniting his interest in
 this problem. We thank Paulo Mour\~ao for initial collaboration in
 this project in 2022, and Alba Grassi and Nils Wagner for useful
 discussions. We also thank Alba Grassi for a critical reading of the
 manuscript.  The work of M.M. was supported in part by the ERC-SyG
 project ``Recursive and Exact New Quantum Theory'' (ReNewQuantum),
 which received funding from the European Research Council (ERC) under
 the European Union's Horizon 2020 research and innovation program,
 grant agreement No. 810573.  The work of J.G. is supported in part by
 the NSF of China through Grant No. 12375062.

\bibliographystyle{JHEP}

\linespread{0.6}
\bibliography{biblio-deformed}

\end{document}